\documentclass[aps,prl,twocolumn,superscriptaddress]{revtex4-1}
\usepackage{amssymb, amsmath, bm}
\usepackage{graphicx,onlyamsmath}

\usepackage{natbib}
\usepackage{xcolor}
\usepackage[normalem]{ulem}  

\begin{document}

\title{Coexistence of Topological and Normal Insulating Phases in Electro-Optically Tuned InAs/GaSb Bilayer Quantum Wells}

\author{M.~Meyer}
\email[]{manuel.meyer@physik.uni-wuerzburg.de}
\affiliation{Julius-Maximilians-Universit\"{a}t W\"{u}rzburg, Physikalisches Institut and W\"{u}rzburg-Dresden Cluster of Excellence ct.qmat, Lehrstuhl f\"{u}r Technische Physik, Am Hubland, 97074 W\"{u}rzburg, Deutschland}

\author{T.~F\"{a}hndrich}
\affiliation{Julius-Maximilians-Universit\"{a}t W\"{u}rzburg, Physikalisches Institut and W\"{u}rzburg-Dresden Cluster of Excellence ct.qmat, Lehrstuhl f\"{u}r Technische Physik, Am Hubland, 97074 W\"{u}rzburg, Deutschland}

\author{S.~Schmid}
\affiliation{Julius-Maximilians-Universit\"{a}t W\"{u}rzburg, Physikalisches Institut and W\"{u}rzburg-Dresden Cluster of Excellence ct.qmat, Lehrstuhl f\"{u}r Technische Physik, Am Hubland, 97074 W\"{u}rzburg, Deutschland}

\author{A.~Wolf}
\affiliation{Julius-Maximilians-Universit\"{a}t W\"{u}rzburg, Physikalisches Institut and W\"{u}rzburg-Dresden Cluster of Excellence ct.qmat, Lehrstuhl f\"{u}r Technische Physik, Am Hubland, 97074 W\"{u}rzburg, Deutschland}

\author{S.~S.~Krishtopenko}
\affiliation{Julius-Maximilians-Universit\"{a}t W\"{u}rzburg, Physikalisches Institut and W\"{u}rzburg-Dresden Cluster of Excellence ct.qmat, Lehrstuhl f\"{u}r Technische Physik, Am Hubland, 97074 W\"{u}rzburg, Deutschland}
\affiliation{Laboratoire Charles Coulomb (L2C), UMR 5221 CNRS-Universit\'{e} de Montpellier, F-34095 Montpellier, France}

\author{B.~Jouault}
\affiliation{Laboratoire Charles Coulomb (L2C), UMR 5221 CNRS-Universit\'{e} de Montpellier, F-34095 Montpellier, France}

\author{G.~Bastard}
\affiliation{Julius-Maximilians-Universit\"{a}t W\"{u}rzburg, Physikalisches Institut and W\"{u}rzburg-Dresden Cluster of Excellence ct.qmat, Lehrstuhl f\"{u}r Technische Physik, Am Hubland, 97074 W\"{u}rzburg, Deutschland}
\affiliation{Physics Department, \'{E}cole Normale Sup\'{e}rieure, PSL 24 rue Lhomond, 75005 Paris, France}

\author{F.~Teppe}
\affiliation{Laboratoire Charles Coulomb (L2C), UMR 5221 CNRS-Universit\'{e} de Montpellier, F-34095 Montpellier, France}

\author{F.~Hartmann}
\email[]{fabian.hartmann@uni-wuerzburg.de}
\affiliation{Julius-Maximilians-Universit\"{a}t W\"{u}rzburg, Physikalisches Institut and W\"{u}rzburg-Dresden Cluster of Excellence ct.qmat, Lehrstuhl f\"{u}r Technische Physik, Am Hubland, 97074 W\"{u}rzburg, Deutschland}

\author{S.~H\"{o}fling}
\affiliation{Julius-Maximilians-Universit\"{a}t W\"{u}rzburg, Physikalisches Institut and W\"{u}rzburg-Dresden Cluster of Excellence ct.qmat, Lehrstuhl f\"{u}r Technische Physik, Am Hubland, 97074 W\"{u}rzburg, Deutschland}

\date{\today}

\begin{abstract}
We report on the coexistence of both normal and topological insulating phases in InAs/GaSb bilayer quantum well induced by the built-in electric field tuned optically and electrically. The emergence of topological and normal insulating phases is assessed based on the evolution of the charge carrier densities, the resistivity dependence of the gap via in-plane magnetic fields and the thermal activation of carriers. For the Hall bar device tuned optically, we observe the fingerprints associated with the presence of only the topological insulating phase. For another Hall bar processed identically but with an additional top gate, the coexistence of normal and topological insulating phases is found by electrical tuning. Our finding paves the way for utilizing a new electro-optical tuning scheme to manipulate InAs/GaSb bilayer quantum wells to obtain trivial-topological insulating interfaces in the bulk rather than at the physical edge of the device.
\end{abstract}

\maketitle

The discovery of topological materials has opened up new avenues in condensed matter physics~\cite{ref1,ref2,ref3,ref4} as topological insulators (TIs) are promising for breakthroughs in fundamental research but also exhibit potential premise for several device applications such as spintronics and quantum computing~\cite{ref5}. Among the vast amount of topological materials~\cite{ref3,ref6,ref7,ref8}, two-dimensional (2D) TIs based on inverted band InAs/GaSb quantum wells (QW) heterostructures~\cite{ref9} are the most appealing for potential device applications due to the mature growth and processing technology developed, which is fully compatible with existing Si-based chips~\cite{ref10} combined with the electrical switching capability of the topological phase transition~\cite{ref9}. The band inversion in InAs/GaSb QW heterostructures is caused by specific band-edge alignment of the InAs and GaSb semiconductors arising at their interface. For thin enough InAs and GaSb layers, the first electron-like (\emph{E}1) level lies above the first hole-like (\emph{H}1) level, and the QW has a trivial band ordering with a normal insulator (NI) phase. As the layer thickness increases, the \emph{E}1 subband becomes lower than the \emph{H}1 subband, which leads to an inverted band structure and the appearance of a 2D TI phase, characterized by an insulating bulk and spin-polarized gapless helical states at the sample edges~\cite{ref9}.

To date, the helical nature of the edge states in InAs/GaSb-based heterostructures has been experimentally reported in InAs/GaSb bilayer quantum wells (BQWs)~\cite{ref11,ref12}. An inherent property of InAs/GaSb BQWs is the lack of inversion symmetry in the growth direction, which directly affects not only the position of the E1 and H1 subbands at the $\Gamma$ point of the Brillouin zone, but also the opening of an inverted band-gap at nonzero quasi-momentum wavevector. Moreover, by changing the strength of the structural asymmetry by applying an external electric field in the growth direction of the BQW, one can change the band ordering in a controlled manner. Therefore, by fabricating a dual-gated device from InAs/GaSb BQWs, it becomes feasible to tune between a normal-insulating (NI) and a TI regime where the helical edge channels are expected~\cite{ref13}. While the electric-field-induced change of the band ordering is often seen as a route towards the realization of a topological field effect transistor~\cite{ref9}, such dual-gating approach also enables another, not yet anticipated, application in topological devices. By depositing several top gates, it becomes possible to create regions with both normal and topological phases in the same sample plane. This should move the interface between a NI- (or vacuum) and 2D TI-phase, which hosts helical edge channels, away from the physical edge of the device into the bulk of the sample.

An improvement of the phase coherence length can then be expected as backscattering via charged donors (or acceptors)~\cite{ref14}, caused by fabrication-related defects, can be significantly reduced. If this strategy is combined with minimizing the residual bulk conductivity (e.g. by substituting the GaSb-layer with GaInSb~\cite{ref15}), major obstacles in the unequivocal observation of the Quantum Spin Hall effect (QSHE) in 2D TIs based on the InAs/GaSb material system~\cite{ref16,ref17,ref18,ref19,ref20} could be overcome. We also emphasize that moving the helical channels away from the sample edge greatly simplifies the fabrication of topological devices, as it allows the helical channels to be guided via relatively simple lithography processes rather than demanding wet chemical etching of nanoscale devices. However, such dual gating of NI and TI interfaces requires not only spatially well-defined top-gate electrodes but also spatially well-defined back-gate electrodes to implement two electric field configurations at a fixed Fermi energy in a single device. To reduce the process challenges of creating numerous gate electrodes, an alternative knob for tuning both electric field and Fermi energy can be used. As shown by Knebl\emph{ et al.}~\cite{ref21}, both the Fermi energy and electric field in InAs/GaSb BQWs can also be tuned optically utilizing a floating gate at the substrate side and negative persistent photoconductivity effect at the surface side~\cite{ref22}. It was shown that by illuminating an InAs/GaSb BQW a top- and back-gating operation can be mimicked, and the Fermi energy can be tuned from an electron-dominated into a hole-dominated regime through the gap with electron-hole hybridization.

In this work, we report the simultaneous observation of normal and topological insulating phases in a single InAs/GaSb BQW due to strong in-plane variations of the built-in electric field across the QW using both electrical and optical tuning methods. For an optically tuned Hall bar device, we observe only a 2D TI phase, whereas for another Hall bar processed identically but with an additional metallic top gate, the coexistence of a NI and 2D TI phase is found. The presence of the NI and 2D TI phases can be assessed by the evolution of the longitudinal resistance with temperature and in-plane magnetic field~\cite{ref13}.

\begin{figure}
\includegraphics [width=1.0\columnwidth, keepaspectratio] {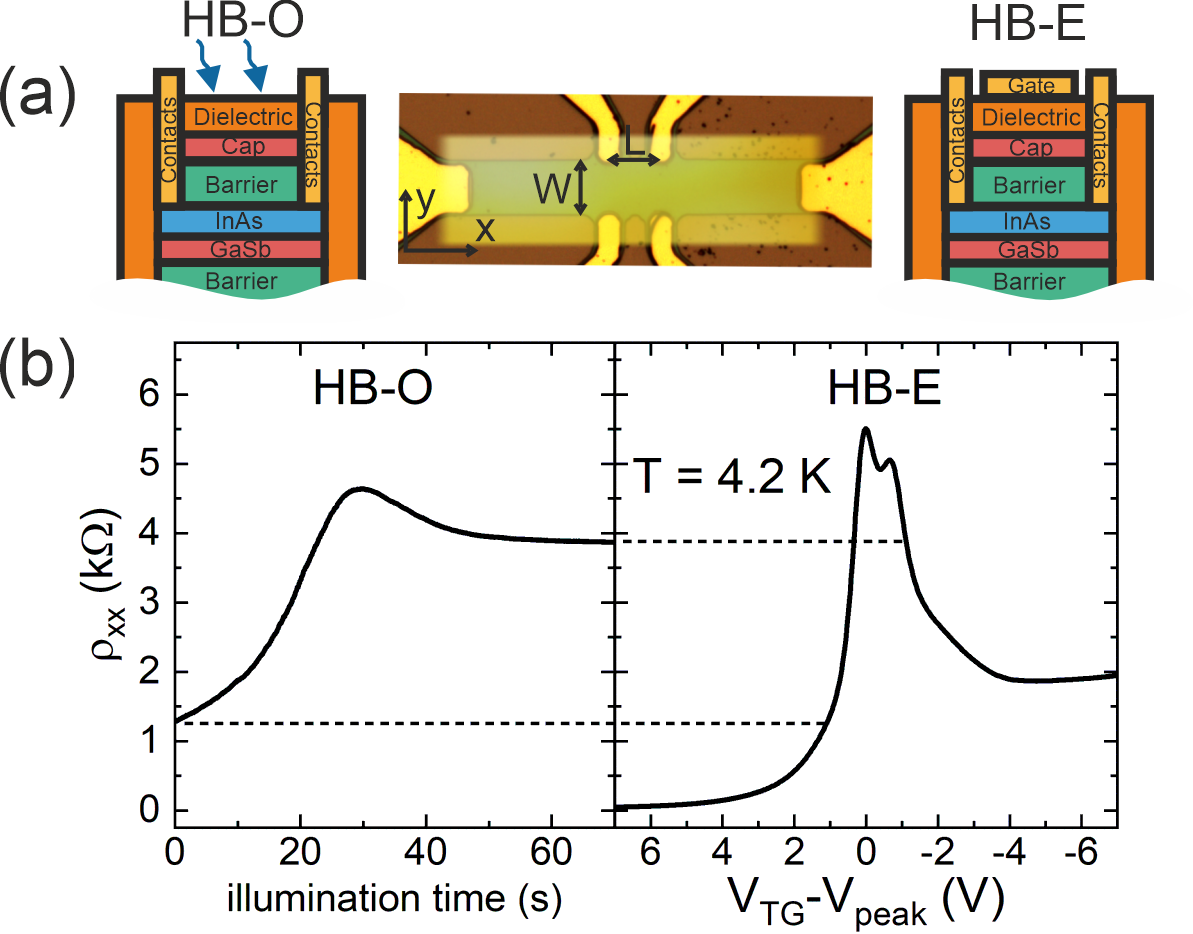} 
\caption{\label{Fig:1} (a) Schematic layout of the devices made from the same InAs/GaSb BQW used for the optical tuning (HB-O, left-hand side) and the electrical tuning (HB-E, right-hand side). Before the deposition of the top-gate electrode, the HB-O device was cleaved from the processing piece. In the middle, an exemplary optical image with a top gate (for HB-E) is displayed. (b) The longitudinal resistivity $\rho_{xx}$ in both devices as a function of the illumination time (left panel) and top-gate voltage (right panel). For HB-E, a unique double-peak structure is observed.}
\end{figure}

Figure~\ref{Fig:1}(a) shows a schematic layer structure of the investigated InAs/GaSb BQW and highlights the difference between the fabricated Hall bar (HB) devices. To compare the transport properties under optical excitation or via a top gate voltage, the HBs were processed from the same wafer and sample piece. The device used for the optical excitation experiments (HB-O) was cleaved from the processing piece just before the metallization of the top gate electrode for the device HB-E. Further information on the device processing is provided in the Supplemental Material~\cite{ref23}. The top-gate electrode is absent for HB-O and instead a UV-LED with an energy of $E_{\mathrm{LED}}\approx5$~eV ($\lambda=248$~nm) at $T=4.2$~K is used. The time scale of the optical tuning depends on the incident light power~\cite{ref21}. Thus, for all measurements, a LED current of $I_{\mathrm{LED}}=2.3$~$\mu$A was used.  Furthermore, an exemplary optical image of a Hall bar with a top gate is shown. Both HBs have the same width $W=20$~$\mu$m but a different length $L=40$~$\mu$m and $60$~$\mu$m for HB-O and HB-E, respectively, where $L$ is the distance between the lateral probes. Due to the different lengths, the resistivity $\rho$ was used in the following analysis. Also labeled are the in-plane directions $x$ (current direction) and $y$.

The studied InAs/GaSb BQW was grown on a (001) GaSb buffer with InAs and GaSb layer thicknesses of $10.5$~nm each. At these layer thicknesses, in the absence of a built-in electric field, band structure calculations~\cite{ref23} indicate that the BQW should be a semimetal with a slight overlap of the conduction and valence band (CB and VB). However, the presence of a built-in electric field, which is always inherent in real samples, opens a band-gap in the band dispersion, turning the semimetal into an insulator. How strong such a built-in electric field needs to be will be addressed later in the manuscript. If the BQW is turned into a TI, the band structure of the BQW is inverted and characterized by a camel-back-like band dispersion of the CB and VB~\cite{ref17}. Note that this a shape of the CB and VB allows for the probing the Van Hove singularities (VHS) associated with the band edges by means of transport measurements~\cite{ref13,ref24,ref25}.

The longitudinal resistivity $\rho_{xx}$ for both devices as a function of the illumination time ($t$) and the top-gate voltage ($V_{TG}$) are presented in Fig.~\ref{Fig:1}(b). If not stated otherwise, all measurements were performed at $T=4.2$~K. Under LED illumination, $\rho_{xx}$ of the HB-O device increases from $\rho_{xx}=1.3$~k$\Omega$ until it reaches its maximum value around $t\approx30$~s with $\rho_{xx,max}=4.6$~k$\Omega$. Subsequently, the longitudinal resistivity drops to $\rho_{xx}=3.9$~k$\Omega$ and remains nearly constant for larger illumination times. As mentioned above, the optical tunability of the HB-O device is attributed to a negative persistent photoconductivity effect (NPPC-effect), which in InAs-based QWs is caused by photo-excited electron-hole pairs in the GaSb cap layer~\cite{ref26,ref27,ref28,ref29} and an accumulation of electrons on the GaSb/AlSb superlattice below the bottom barrier~\cite{ref21}. Note that the illumination effects are persistent but can be reverted by switching off the LED and heating up the sample. More detailed information on the optical tunability of InAs/GaSb BQWs is given in Ref.~\cite{ref21}.

In contrast, the longitudinal resistivity of HB-E is tuned with a top gate. The central resistivity peak at $V_{TG}=V_{peak}$ occurs when the Fermi level lies inside the band-gap of the InAs/GaSb BQW. For positive values of $V_{TG}-V_{peak}$, the Fermi level lies in the CB, while decreasing of $\rho_{xx}$ at negative $V_{TG}-V_{peak}$ is caused by the Fermi level shifting into the VB. As seen, in the vicinity of its maximum value, $\rho_{xx}$ shows a unique double-peak structure. Although similar double-peak features in InAs/GaSb BQWs were also reported in other works~\cite{ref13,ref24}, their origin still remains unclear as they were not further investigated. Here, the $\rho_{xx}$ peak values are $5.5$~k$\Omega$ and $5.1$~k$\Omega$ for the left and right peak, respectively. The dashed lines contrast the range of the optical compared to the electrical tuning. As can be seen, the full illumination time used for the optical tuning corresponds to the respective gate voltage range of about ${\Delta}V_{TG}\approx2.5$~V. Although the optical tuning range of $\rho_{xx}$ is smaller than for the electrical gating, it is still possible to tune the Fermi level position from the CB to the VB through the gap region where the helical edge states appear. However, the maximum resistance value is also roughly 20\% smaller in the latter case than in the HB-E device. In the following, the reasons for the difference in both devices will be explored.

\begin{figure}
\includegraphics [width=1.08\columnwidth, keepaspectratio] {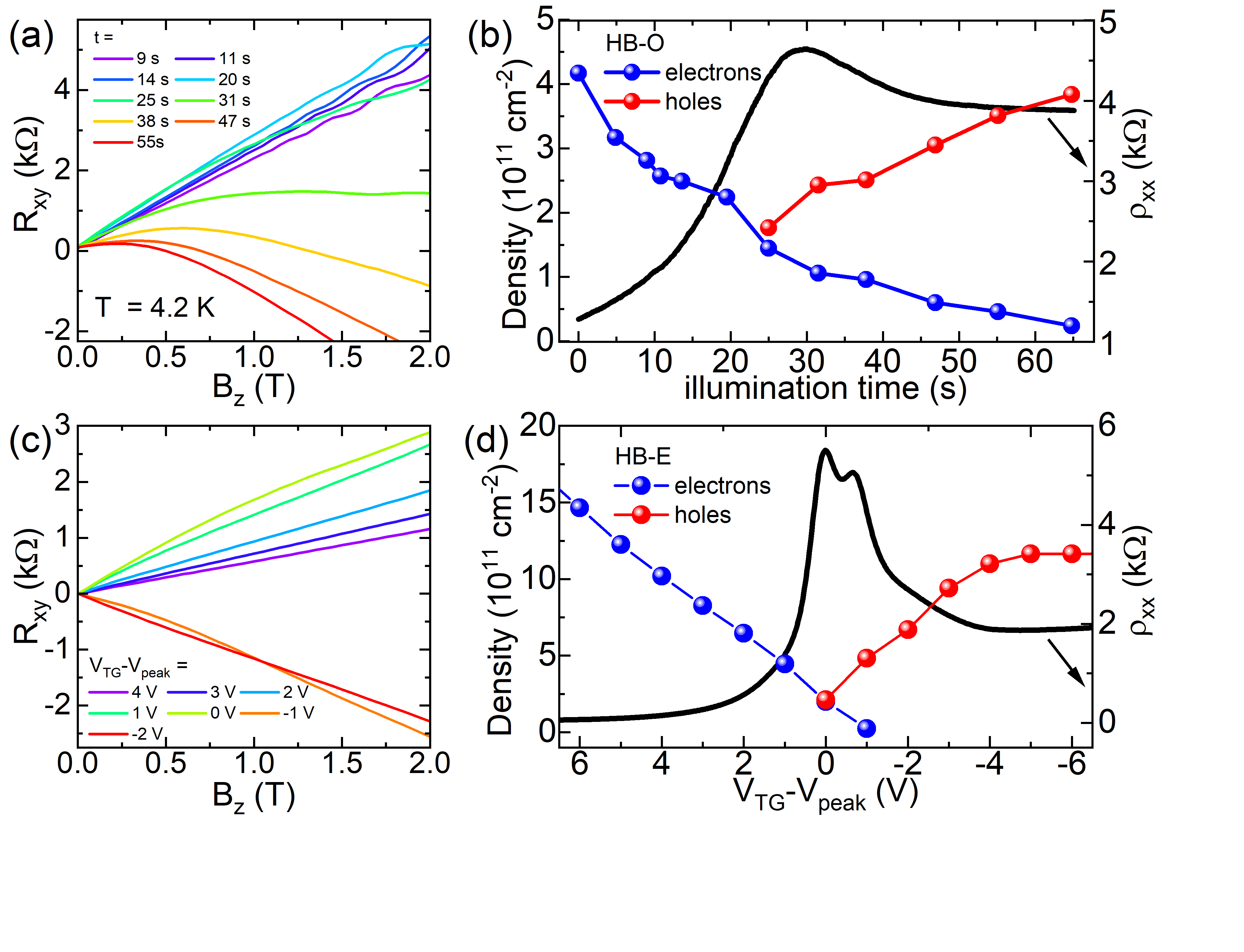} 
\caption{\label{Fig:2} (a) Hall resistance traces for different illumination times ranging from $t=9$~s up to $55$~s. Starting from $t=25$~s, a pronounced nonlinearity is observed indicating two-carrier transport. (b) Extracted electron (in blue) and hole (in red) densities and the zero-field resistivity $\rho_{xx}$ (in black) for different illumination times. After $t=25$~s, an electron and hole densities can be extracted indicating electron-hole hybridization. (c) Hall resistance $R_{xy}$ versus magnetic field for $V_{TG}-V_{peak}=+4$~V to $-2$~V. In contrast to HB-O, $R_{xy}$ does not show pronounced nonlinearities. (d) Electron (in blue) and hole (in red) charge carrier densities and $\rho_{xx}$ at zero magnetic field (in black) as a function of top-gate voltage normalized to the central peak $V_{peak}$. Within a small region around the gap, both carrier types are present.}
\end{figure}

To follow the evolution of the charge carrier densities, magneto-transport measurements for HB-O and HB-E with an out-of-plane magnetic field were performed. In Fig.~\ref{Fig:2}(a), the Hall measurements for HB-O for different illumination times $t=9-55$~s and $B_{z}=0-2$~T are presented. For short illumination times, the Hall resistance ($R_{xy}$) is linear, indicating that only electrons are present and Quantum Hall plateaus are visible. However, at larger illumination times, a pronounced nonlinearity is observed, which is attributed to the presence of two distinct charge carriers. An analysis of $R_{xy}$ in a magnetic field, performed on the basis of a two-charge-carrier model~\cite{ref30}, provides the electron (blue) and hole (red) densities as a function of the illumination time, summarized in Fig.~\ref{Fig:2}(b) together with the zero-field resistivity $\rho_{xx}$ (in black). At shorter illumination times, electrons are the majority charge carriers with a maximum density of $n_{max}=4.2\cdot10^{11}$~cm$^{-2}$ and mobility $\mu_{max}=2.6\cdot10^4$~cm$^2$/V$\cdot$s. Starting from $t=25$~s, both electron and hole densities are observable. As they coincide, they represent a charge neutrality point (CNP) in the system. Further increasing t switches the charge-carrier transport gradually from \emph{n-} to \emph{p-}type. Thus, the HB-O device features a broad range of illumination time with an electron-hole hybridization regime, which is an inherent property of a 2D TI phase in InAs/GaSb BQWs~\cite{ref13}.

Similar measurements were performed for the HB-E device. Figure~\ref{Fig:2}(c) displays the Hall resistance as a function of magnetic field $B_{z}=0-2$~T shown for a chosen range of $V_{TG}-V_{peak}=+4$~V to $-2$~V. In contrast to the HB-O device, $R_{xy}$ remains linear for almost all gate voltage values except for $V_{TG}-V_{peak}=0$ and $-1$~V, a weak nonlinearity is observable in $R_{xy}$. In addition, no quantum Hall plateaus are visible in the low magnetic field range due to the electron mobility being roughly 50\% lower compared to HB-O device. The results of the analysis of $R_{xy}$ within the two-charge-carrier model are shown in Fig.~\ref{Fig:2}(d). For positive values of $V_{TG}-V_{peak}$, only electrons are present with the highest mobility $\mu_{max}=1.6\cdot10^5$ cm$^2$/V$\cdot$s at $n_{max}=2.7\cdot10^{12}$~cm$^{-2}$. This mobility value is roughly a factor 4 lower than the highest values reported for BQWs~\cite{ref31,ref32}. However, it should still be sufficient for observing both VHS in the CB and VB using solely the top gate if the sample is in the TI regime. Around the maximum value of $\rho_{xx}$ there is a small range of the gate voltages $V_{TG}-V_{peak}=0$ to $-1$~V, in which electrons and holes coexist. However, the range of the electron-hole hybridization does not extend into the VB as it is expected for a 2D TI phase in InAs/GaSb BQWs~\cite{ref13}. At negative $V_{TG}-V_{peak}$ values, less than $-1$~V, only holes remain as charge carriers, whose maximum mobility reaches $\mu_{max}=6\cdot10^3$ cm$^2$/V$\cdot$s at $p_{max}=1.2\cdot10^{12}$cm$^{-2}$.

Despite the qualitatively similar dependencies of the longitudinal resistivity $\rho_{xx}$ in the HB-O and HB-E devices, the distinctive ranges for the electron-hole hybridization may indicate different topological phases of the InAs/GaSb BQW in both devices~\cite{ref13}. In-plane magnetic field dependent measurements are a powerful tool to distinguish between a NI and TI gap in InAs/GaSb BQWs. An in-plane magnetic field $B_{||}$ along the $x$-direction shifts the bands in the $k_y$-direction and vice versa by ${\Delta}k_{x/y}=eB_{y/x}\langle{z}\rangle/\hbar$, where $\langle{z}\rangle$ is the average distance between the electron and hole gases~\cite{ref33,ref34}. This leads to a phase transition into the semi-metallic phase for a TI sample, whereas a NI sample is almost unaffected~\cite{ref13,ref35a}.

\begin{figure}
\includegraphics [width=1.6\columnwidth, keepaspectratio] {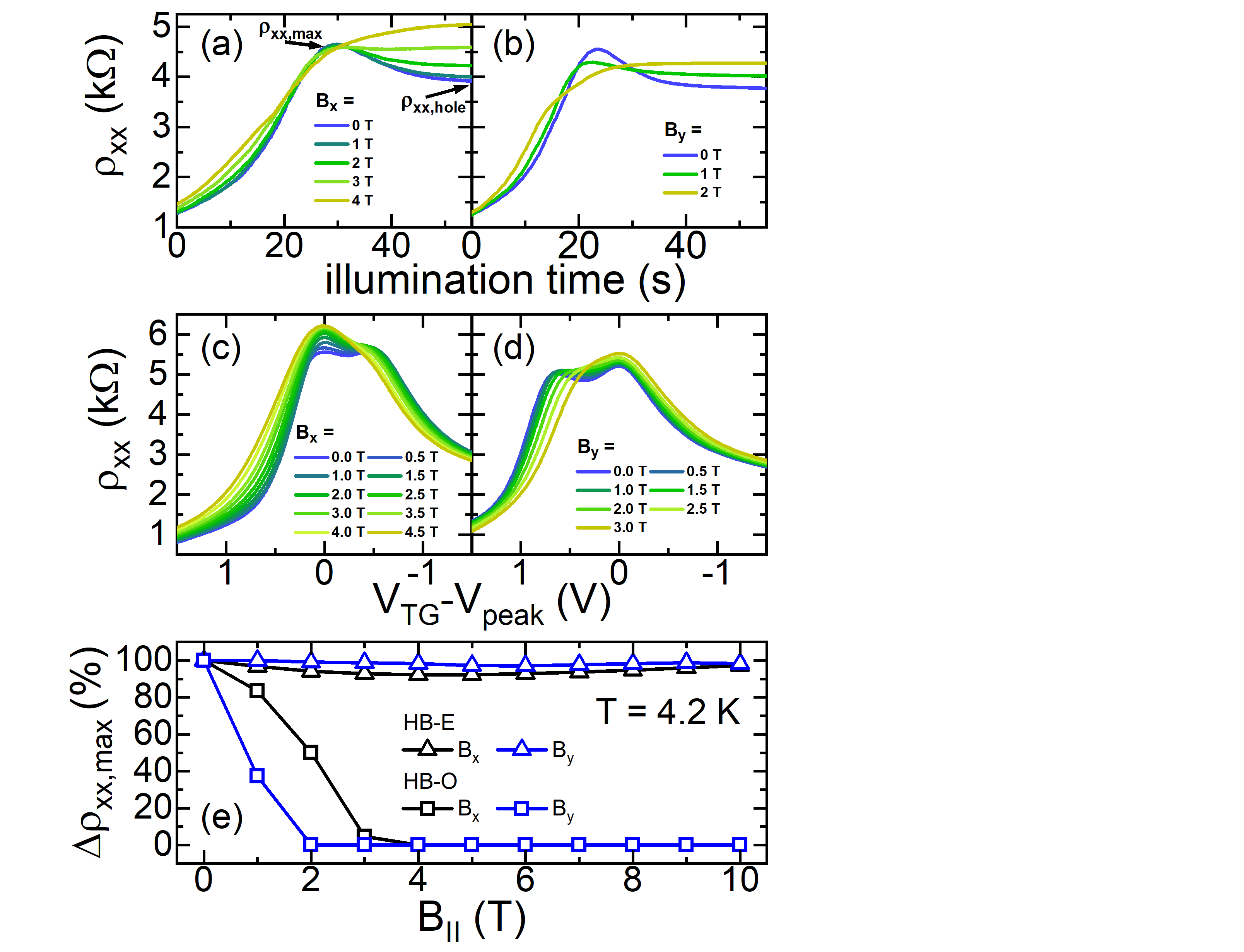} 
\caption{\label{Fig:3} (a,b) $\rho_{xx}$ as a function of illumination time for different in-plane magnetic fields $B_x$ (panel (a)) and $B_y$ (panel (b)). For $B_x=4$~T and $B_y=2$~T, the resistivity peak vanishes indicating the closing of the TI gap. (c,d) $\rho_{xx}$ versus $V_{TG}-V_{peak}$ for different in-plane magnetic fields $B_x$ (panel (c)) and $B_y$ (panel (d)). The double-peak feature vanishes at around $B_x=4.5$~T and $B_y=2.5$~T, and the remaining peak prevails until 10~T for both in-plane magnetic field orientations. (e) The relative change in the difference in the longitudinal resistance between the peak and saturated value at the VB $\Delta\rho_{xx,max}$ as a function of the in-plane magnetic field. For the HB-O device, $\Delta\rho_{xx,max}$ drops to zero, while it remains close to unity in the HB-E device.}
\end{figure}

Figure~\ref{Fig:3}(a) and (b) show the evolution of $\rho_{xx}$ versus $B_x$ and $B_y$, respectively. Please note that or the highest applied in-plane magnetic field of 10~T, a small $z$-component of a maximum of 100 mT~could be extracted for both devices, which is negligible. The resistivity peak vanishes with increasing magnetic fields, indicating the band-gap closing and transition to a semi-metallic state at $B_x=4$~T and $B_y=2$~T. We note that the anisotropy of the evolution of $\rho_{xx}$ with respect to different orientations of the in-plane magnetic field is similar to the one observed in Ref.~\cite{ref13}. In contrast, only one of the $\rho_{xx}$ peaks in the HB-E device vanishes with increasing in-plane magnetic field, as shown in Fig.~\ref{Fig:3}(c) and (d). This occurs at around $B_x=2.5$~T and $B_y=4.5$~T, while the other peak survives up to $B_{x/y}=10$~T -- the highest field available in our experiments. We emphasize that the fields, at which this resistivity peak disappears, are comparable to those at which the peak vanishes in the HB-O device. This indicates that the vanishing $\rho_{xx}$ peak in the HB-E device may also be associated with a gap closing of a 2D TI phase. On the other hand, the other resistivity peak immune to the in-plane magnetic field, clearly indicates the presence of a NI phase in the InAs/GaSb BQW of the HB-E device~\cite{ref13}. Thus, the HB-E device features the fingerprints of both NI and 2D TI phases in the InAs/GaSb BQW. For better visualization, Fig.~\ref{Fig:3}(e) illustrates the band-gap evolution in both devices by providing the relative change in the difference in $\rho_{xx}$ between the maximum value $\rho_{xx,max}$ and saturated value in the VB $\rho_{xx,hole}$ as a function of the in-plane magnetic field:
$\Delta\rho_{xx,max}(B_{||})=\{\rho_{xx,max}(B_{||})-\rho_{xx,max}(B_{||})\}/\{\rho_{xx,max}(0)-\rho_{xx,max}(0)\}$. As clearly seen, $\Delta\rho_{xx,max}(B_{||})$ of the HB-O device drops to zero with increasing in-plane magnetic field, while it remains close to unity for the HB-E device.

\begin{figure}
\includegraphics [width=1.28\columnwidth, keepaspectratio] {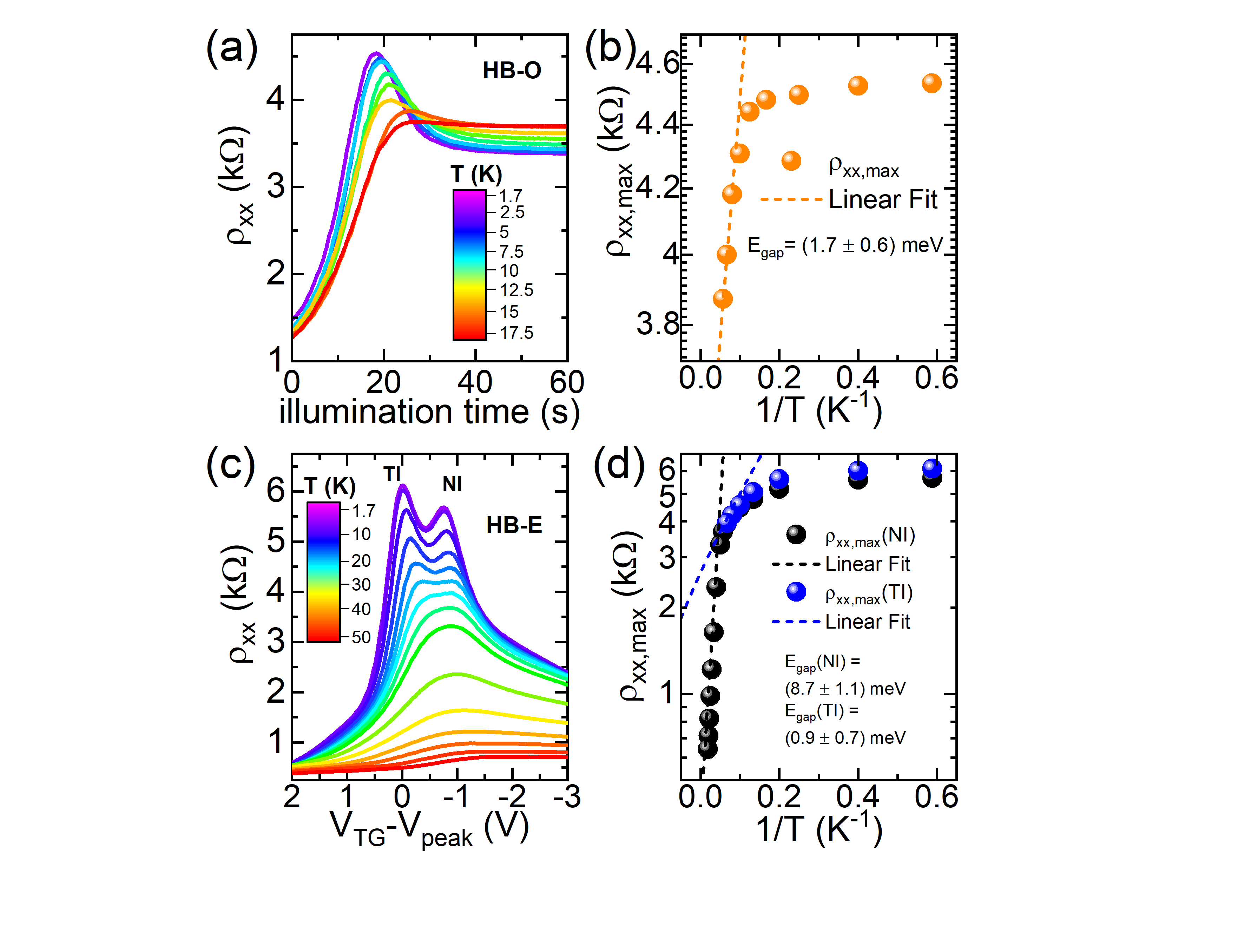} 
\caption{\label{Fig:4} (a) $\rho_{xx}$ versus illumination time for different temperatures from $T=1.7$~K up to $17.5$~K for HB-O device. The resistance peak vanishes at $T=17.5$~K. (b) Arrhenius plot of $\rho_{xx,max}(1/T)$. The fitting at high temperatures yields a band-gap $E_{gap}=(1.7\pm0.6)$~meV. (c) $\rho_{xx}$ versus illumination time for different temperatures from $T=1.7$ to $50$~K for HB-E device. The left-hand peak for the TI phase and the right-hand peak for the NI phase vanish around $T=15$~K and $50$~K, respectively. (d) Arrhenius plot of the left-hand (in blue) and right-hand (in black) peak values $\rho_{xx,max}$. The fit of the high-temperature regime gives two band-gap values: $E_{gap}(TI)=(0.9\pm0.7)$~meV and $E_{gap}(NI)=(8.7\pm1.1)$~meV, attributed to 2D TI and NI phase, respectively.}
\end{figure}

To determine the band-gap values associated with each of the longitudinal resistivity peaks, temperature-dependent measurements have also been performed for both devices (see Fig. 4). For HB-O device in Fig.~\ref{Fig:4}(a), the peak corresponding to the gap vanishes around $T=17.5$~K. Figure~\ref{Fig:4}(b) shows the Arrhenius plot of the peak resistivity in the HB-O device and its high-temperature fit by the activation contribution $\exp(-E_{gap}/2k_{B}T)$, where $k_{B}$ is the Boltzmann constant~\cite{ref15,ref21}. The fit gives the band-gap energy, $E_{gap}=(1.7\pm0.6)$~meV. At low temperatures, the dependence is weak which might indicate hopping transport~\cite{ref35}. The same analysis has been performed for the temperature dependence of each of the resistivity peaks for the HB-E device (see Fig.~\ref{Fig:4}(c)) and the Arrhenius plot is depicted in Fig.~\ref{Fig:4}(d). Contrary to HB-O device, the double peak for HB-E device allows for the extraction of two band-gap energies. One peak vanishes around $T=15$~K (left-hand peak, TI) while the second peak persists until $T=50$~K (right-hand peak, NI). Two different band-gap energies can be determined: $E_{gap}(TI)=(0.9\pm0.7)$~meV and $E_{gap}(NI)=(8.7\pm1.1)$~meV. The former value coincides within the error bar with the extracted value for HB-O device.  Furthermore, the left-hand side peak shows the same evolution with the in-plane magnetic field as the $\rho_{xx}$-peak in the HB-O device. Therefore, we attribute the left-hand peak to the 2D TI phase, as it also occurs at higher VTG-values similar to Ref.~\cite{ref13}. In contrast, the higher value for the right-hand peak is attributed to another phase as it significantly exceeds the maximum gap energy in the TI regime~\cite{ref13,ref24,ref33} and also does not vanish with in-plane magnetic fields. Hence, the right peak and the associated band-gap is attributed to a NI phase. We conclude that these measurements provide evidence that a mixed NI/TI phase is achieved by tuning the BQW solely with a top gate.

The key to understanding these different results obtained for two devices is implicitly presented in Figs~\ref{Fig:2}(b) and \ref{Fig:2}(d). As is clear from Fig.~\ref{Fig:2}(b), a 2D TI phase, whose existence has been unequivocally proven experimentally in the HB-O device, is characterized by the presence of both electrons and holes originating from the camelback shape of the bands. Since the electron-like \emph{E}1 subband and the hole-like \emph{H}1 subband are localized in the InAs- and GaSb-layer, respectively~\cite{ref9}, the presence of electrons and holes results in a charge carrier separation in the QW, hence the appearance of a built-in electric field. In real samples, the built-in field is determined not only by spatially separated electrons and holes, but also by the distribution of charged donors and acceptors in the layers external to the BQW. However, just from the difference in the electron and hole concentrations in Figs~\ref{Fig:2}(b) and \ref{Fig:2}(d), one can already conclude that the built-in electric field in both devices should differ dramatically. Since the electric field may yield a transition between the semimetal, 2D TI and NI phases in InAs/GaSb BQW, the difference in the fields may explain the observation of different phases in two devices made from nominally the same QW.

\begin{figure}
\includegraphics [width=1.04\columnwidth, keepaspectratio] {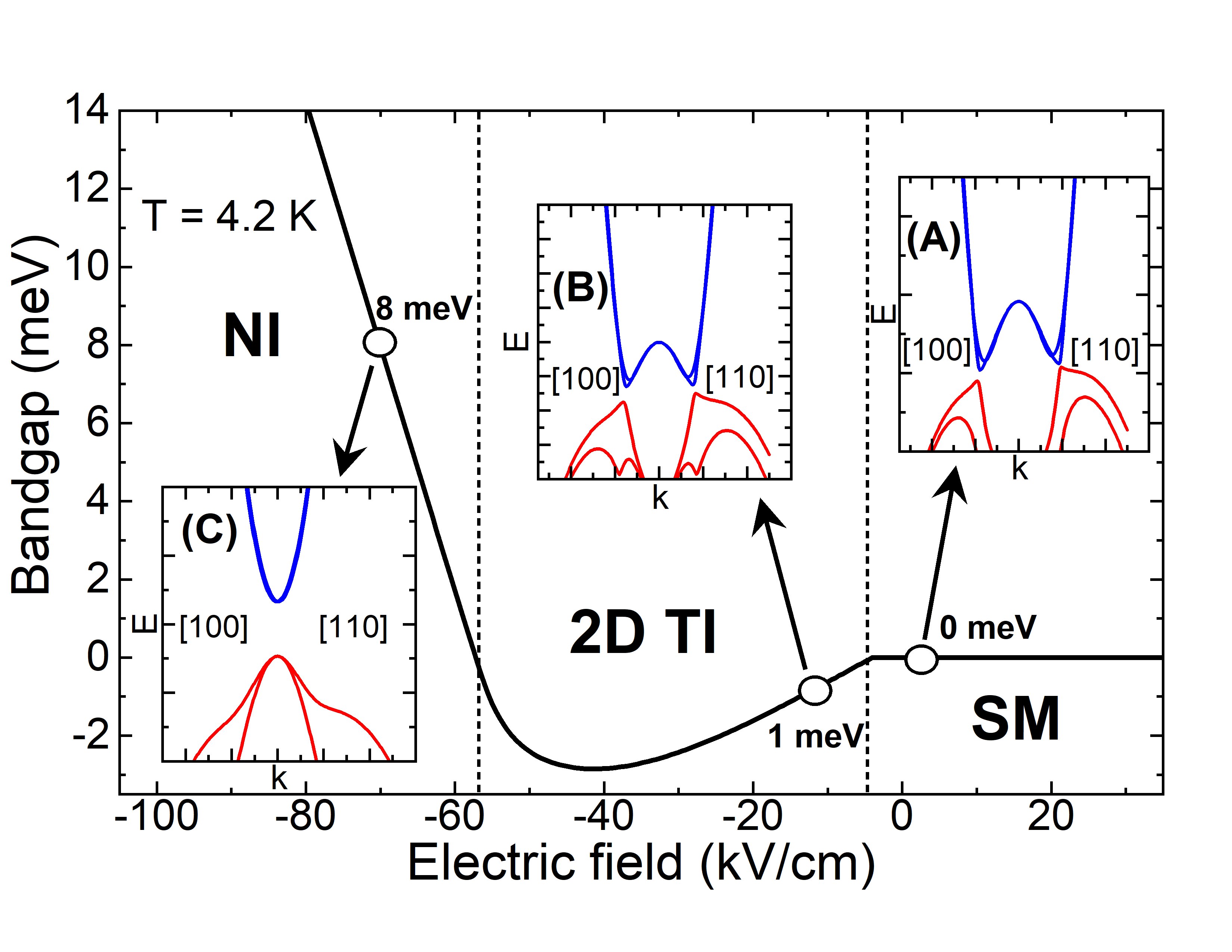} 
\caption{\label{Fig:5} Band-gap of the InAs/GaSb BQW of the HB-O and HB-E devices as a function of the effective electric field along the growth direction. Positive and negative band-gap values correspond to NI and 2D TI phase, respectively. Negative electric field values correspond to the orientation of the electric field strength in the direction from the substrate to the surface (see Fig.~\ref{Fig:1}). The insets show a plot of band dispersion for several electric field values: (A)~no~electric field, semimetal (SM) phase; (B)~$-13.5$~kV/cm, 2D TI phase with $E_{gap}(TI)=1$~meV and (C)~$-70$~kV/cm, NI phase with $E_{gap}(NI)=8$~meV. For the insets, the separation of the main vertical ticks is $10$~meV and for the horizontal $0.1$~nm$^{-1}$.}
\end{figure}

Since the band-gap in InAs/GaSb BQW depends on the electric field~\cite{ref9}, one can estimate the strength and orientation of the \emph{effective} built-in electric field across the QW in each device by using experimental values obtained from temperature-dependent transport measurements (see Fig.~\ref{Fig:4}). Figure~\ref{Fig:5} shows the band-gap of the InAs/GaSb BQW of the HB-O and HB-E devices at T = 4.2 K as a function of the effective electric field across the QW~\cite{ref23,ref36}. The positive orientation of the electric field corresponds to the direction from the GaSb cap layer to the substrate (see a schematic layout in Fig.~\ref{Fig:1}(a)). As seen, the application of a positive electric field to the InAs/GaSb BQW, which is a semimetal in the absence of an electric field, only increases the overlap between the CB and VB. On the contrary, a negative field first opens the band-gap in the InAs/GaSb BQW, causing a sequential transition from the semimetal into the 2D TI phase, and then induces a second phase transition from the 2D TI to the NI phase at high field values. Typical 2D plots of band dispersion for the semimetal, 2D TI and NI phases tuned by electric field are provided in the insets of Fig.~\ref{Fig:5}. Without an electric field, the sample should be semi-metallic (panel (A)). The 2D TI phase with a band-gap of 1 meV (a value that fits well within the error bars of the experimental values obtained in two devices at once) is realized at an effective electric field of about $-13.5$~kV/cm (panel (B)). This value seems very reasonable since it is significantly less than the electric field value corresponding to the maximally asymmetric profile of the distribution of charge carrier suppliers in ungated InAs/AlSb-based QWs~\cite{ref27,ref37}.

The NI phase with a band-gap of about 8~meV, found in the HB-E device, corresponds to a much larger \emph{effective} electric field with a strength of about $-70$~kV/cm (panel (C)). At this point, it becomes apparent that some inhomogeneities are involved that cause the two different phases. One possibility would be inhomogeneities in the sample, such as layer fluctuations. However, these would also be present in HB-O device, where only one phase is observable. The second and most likely possibility is that when tuning with the top gate, there are two different electric field configurations randomly distributed over the area of the Hall bar. These cause the occurrence of NI and TI phases over the area of the Hall bar. From a simple resistor network model (see Supplemental Material~\cite{ref23}) it is possible to estimate the ratio of the area coverage of both phases for HB-E device. Using this model, we found that approximately 80\% of the Hall bar is covered by a TI phase and 20\% by a NI phase.

In summary, we have investigated NI and 2D TI phases arising in InAs/GaSb BQWs due to the built-in electric field across the QW, which can be tuned optically and electrically. For the Hall bar device tuned optically, we have clearly observed the presence of only a 2D TI phase, while for another Hall bar processed identically but tuned with an additional top gate, the coexistence of a NI and 2D TI phase has been found. The differences between the NI and 2D TI phase have been experimentally discriminated by means of the measured evolution of longitudinal resistance versus temperature and in-plane magnetic field. These experimental observations show the possibility of an electro-optical controlled phase diagram, which results in further device flexibility. With the finding of a mixed NI/TI phase, it becomes possible to distance the helical edge channels from the processed physical edge of the sample by creating an NI/TI interface in the bulk. Therefore, the helical edge channels can be separated from the trivial edge channels, which would make it more feasible to observe the Quantum Spin Hall effect in this material system.

\begin{acknowledgments}
The work was supported by the Elite Network of Bavaria within the graduate program ``Topological Insulators'' and by the French Agence Nationale pour la Recherche (ANR) with ``Equipex+ Hybat'' (ANR-21-ESRE-0026) project. We acknowledge financial support from the DFG through the W\"{u}rzburg-Dresden Cluster of Excellence on Complexity and Topology in Quantum Matter -- ct.qmat (EXC 2147, project-id 390858490) and by the Occitanie region through the program ``Quantum Technologies Key Challenge'' (TARFEP project).
\end{acknowledgments}


%


\newpage
\clearpage
\setcounter{equation}{0}
\setcounter{figure}{0}
\setcounter{table}{0}
\renewcommand{\thefigure}{S\arabic{figure}} %
\renewcommand{\thetable}{S\arabic{table}}   %
\renewcommand{\theequation}{S\arabic{equation}}   %

\newpage
\clearpage
\setcounter{equation}{0}
\setcounter{figure}{0}
\setcounter{table}{0}
\renewcommand{\thefigure}{S\arabic{figure}} %
\renewcommand{\thetable}{S\arabic{table}}   %
\renewcommand{\theequation}{S\arabic{equation}}   %

\onecolumngrid
\begin{center}
\LARGE{\textbf{Supplemental Material}}
\end{center}
\maketitle
\onecolumngrid

\subsection{A. Sample growth and processing}
\begin{figure}[h!]
\includegraphics [width=0.5\columnwidth, keepaspectratio] {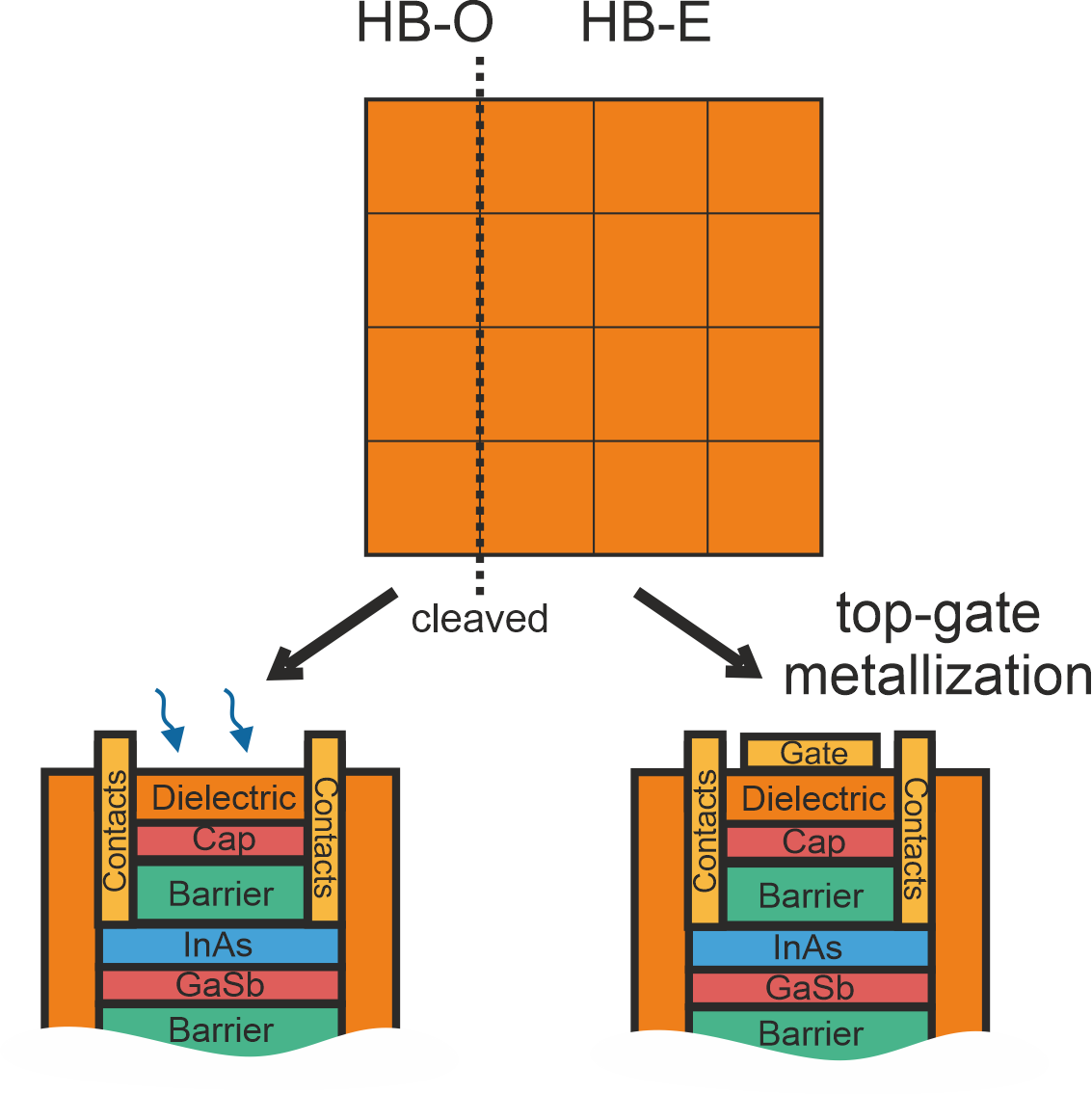} 
\caption{\label{Fig:S1} Fabrication process of the Hall bars for both devices. They were manufactured on the same wafer piece and only after the gate dielectric was deposited the piece was cleaved. Afterwards, one column of the Hall bars was used for the fabrication of the HB-O devices intended for the optical tuning, while on the other columns, the top-gate electrodes were deposited on top of the Hall bars (the HB-E device).}
\end{figure}

The sample was grown by MBE on an undoped (001) GaSb-substrate followed by a 2000~nm thick undoped buffer layer. On top of the buffer is a $10\times2.5$/$2.5$~nm GaSb/AlSb superlattice (SL). After the SL, the InAs/GaSb bilayer quantum well (BQW) (10.5/10.5~nm) is sandwiched between a bottom 100 nm thick AlAs$_{0.08}$Sb$_{0.92}$-barrier and the top 50~nm thick AlAs$_{0.08}$Sb$_{0.92}$-barrier. The sample is capped with a $5$~nm thick GaSb-layer. After the growth, a gate dielectric consisting of $5\times10$/$10$~nm SiO$_2$/Si$_3$N$_4$ was deposited. After the deposition, the sample piece was cleaved as shown in Fig.~\ref{Fig:S1}. The first column of Hall bars was used for the fabrication of the HB-O devices tuned optically with no metallic top gate deposited. For the HB-E device, a $30$~nm Cr and $100$~nm Au top gate was deposited on top. Therefore, the investigated devices are nominally identical.

\subsection{B. Band structure calculations}
\begin{figure*}
\includegraphics [width=0.5\columnwidth, keepaspectratio] {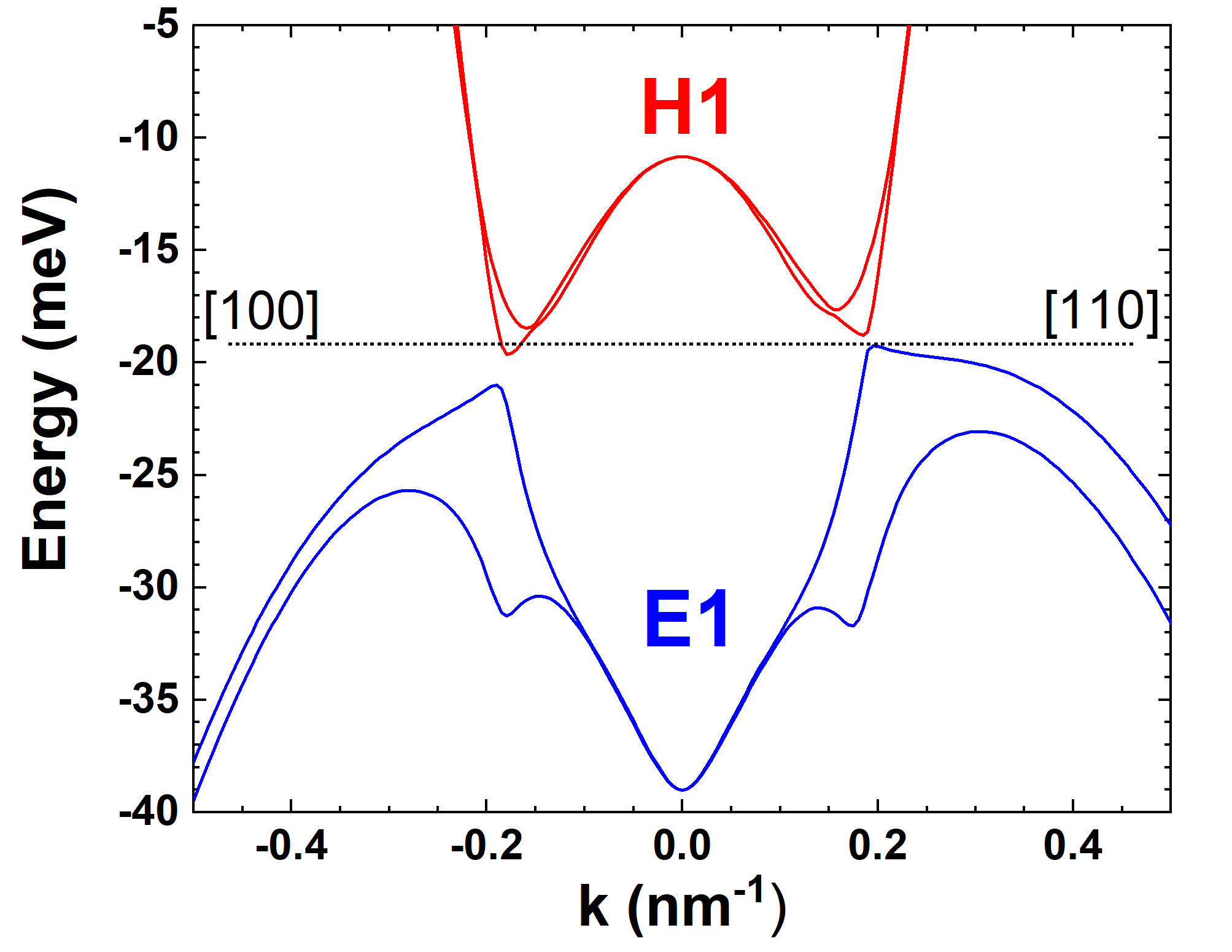} 
\caption{\label{Fig:S2} Band structure of the InAs/GaSb BQW studied in this work, based on an eight-band \textbf{k$\cdot$p} Hamiltonian in the absence of built-in electric field across the QW. The blue and red curves represent the energy-dispersion of the electron-like (\emph{E}1) and hole-like (\emph{H}1) subbands, respectively. The positive and negative values of quasimomentum k correspond to the [100] and [110] crystallographic orientations. The horizontal dashed line highlights the slight overlap between the bottom of the conduction band in the [100] direction and the top of the valence band in the [110] direction. The absence of an inversion center in the growth direction of InAs/GaSb BQW leads to the Kramers spin degeneracy lifting at non-zero wave-vector, known as Rashba spin-splitting.}
\end{figure*}

Band structure calculations in the main text have been performed by using the eight-band \textbf{k$\cdot$p} Hamiltonian~\cite{SMref1}, which directly takes into account the interactions between $\Gamma_6$, $\Gamma_8$ and $\Gamma_7$ bands in bulk materials. This model describes well the electronic states in a wide range of narrow-gap semiconductor QWs including InAs/GaSb BQWs~\cite{SMref2}. In the Hamiltonian, we also consider the terms describing the strain effect arising due to mismatch of lattice constants in the buffer, QW layers and AlSb barriers. The calculations have been performed by expanding the eight-component envelope wave functions in the basis set of plane waves and by numerical solution of the eigenvalue problem. Details of calculations, the form of the Hamiltonian can be found elsewhere~\cite{SMref1}. Parameters for the bulk materials, and valence band offsets used in the calculations are taken from Ref.~\cite{SMref3}. Figure~\ref{Fig:S2} shows the band structure calculations for the sample under study in the absence of built-in electric field across the QW. The results in the presence of effective electric field are provided in Fig.~5 of the main text.

\begin{figure}[h!]
\includegraphics [width=0.45\columnwidth, keepaspectratio] {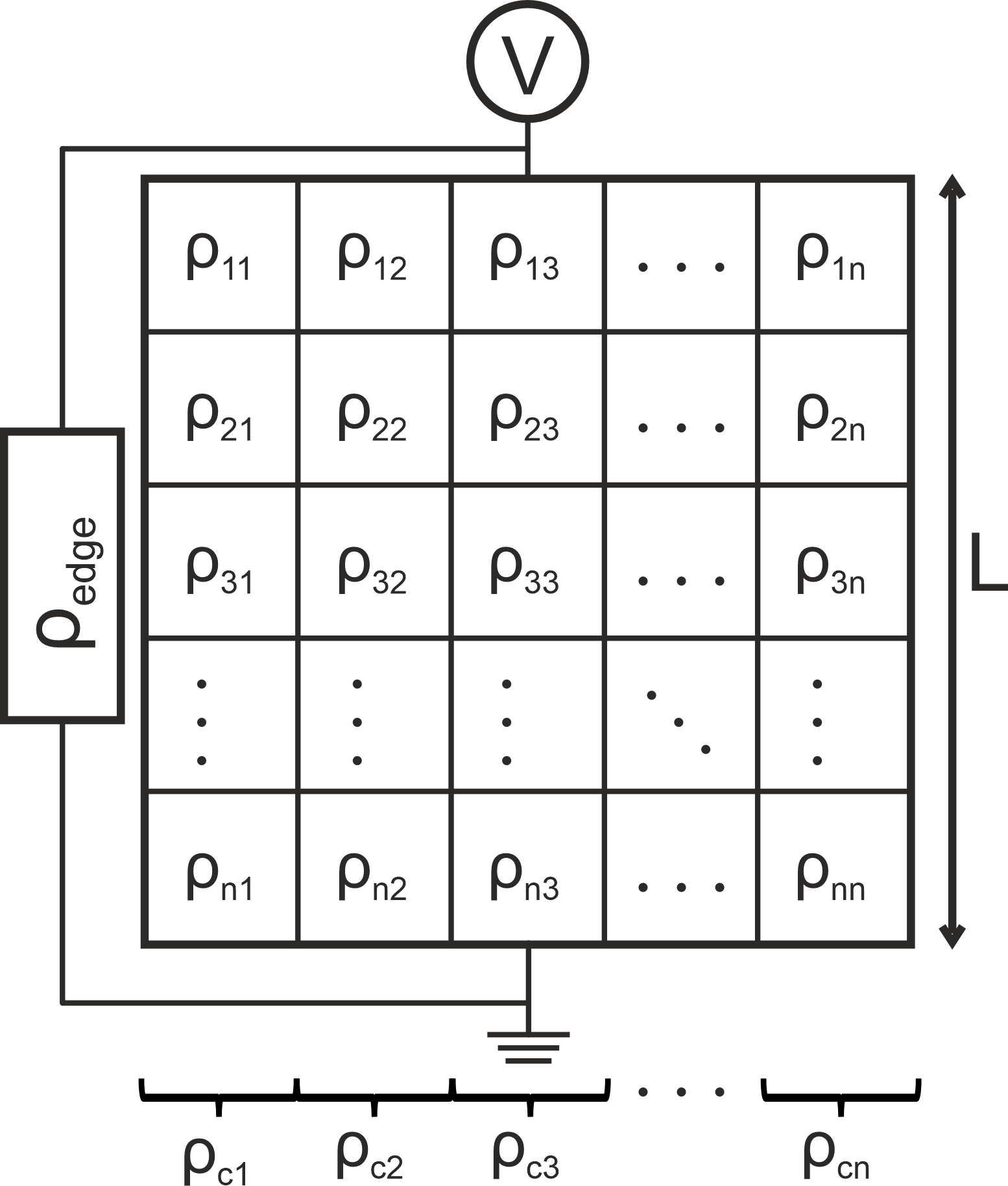} 
\caption{\label{Fig:S3} Simplified resistor network with $n{\times}n$ building blocks to model the double-peak shape observed for HB-device.}
\end{figure}

\subsection{C. Resistor network model}
To describe the shape of the double peak we have applied a resistor network model~\cite{SMref4,SMref5}, which was simplified as horizontal transport is neglected. A $n{\times}n$ square is used and the voltage $V$ is applied from top to bottom as shown in Fig.~\ref{Fig:S3}. Each of the building blocks has a resistance labeled from $\rho_{11}$ to $\rho_{nn}$. Therefore, the resistance for each column are just the resistances of the building blocks added and as an example for $\rho_{c1}$:
\begin{equation}
\label{eq:C1}
\rho_{c1}=\rho_{11}+\rho_{21}+\rho_{31}+...+\rho_{n1}.
\end{equation}
For the total resistance for the bulk, it follows:
\begin{equation}
\label{eq:C2}
\rho_{bulk}=\left(\dfrac{1}{\rho_{c1}}+\dfrac{1}{\rho_{c2}}+\dfrac{1}{\rho_{c3}}+...+\dfrac{1}{\rho_{cn}}\right)^{-1}.
\end{equation}
One could also take the helical ($\rho_{helical}$) and trivial ($\rho_{trivial}$) edge channels into account. As the helical edge channels normally only have a length of a few $\mu$m, they do not need to be considered for the macroscopic devices. For the trivial edge channels in a InAs/GaSb BQW, they often possess a resistivity of a few k$\Omega$/$\mu$m (e.g. 2 k$\Omega$/$\mu$m~\cite{SMref6}). In this model, for $L>10$~$\mu$m, the trivial edge channels do not change the shape of the double peak anymore and are therefore negligible as well. Then, the total resistance $\rho_{total}$ can be described as:
\begin{equation}
\label{eq:C3}
\rho_{total}=\left(\dfrac{1}{\rho_{bulk}}+\dfrac{1}{\rho_{helical}}+\dfrac{1}{\rho_{trivial}}\right)^{-1}\simeq\rho_{bulk}.
\end{equation}

Now, we assume that each building block can either be in a NI or TI phase, where the complete resistance is labeled as $\rho_{NI}$ or $\rho_{TI}$, respectively. For each phase, the peak for the gap can be approximated by a Gaussian function:
\begin{equation}
\label{eq:C4}
\rho_{NI}=\rho_{max,NI}\times\exp\left(-\dfrac{\left(V-V_{NI}\right)^2}{S_{NI}}\right);~~~~~~~~
\rho_{TI}=\rho_{max,TI}\times\exp\left(-\dfrac{\left(V-V_{TI}\right)^2}{S_{TI}}\right),
\end{equation}
where $\rho_{max,NI}$ and $\rho_{max,TI}$ are the respective maximum resistance values for the band-gap.

\begin{figure}[h!]
\includegraphics [width=0.65\columnwidth, keepaspectratio] {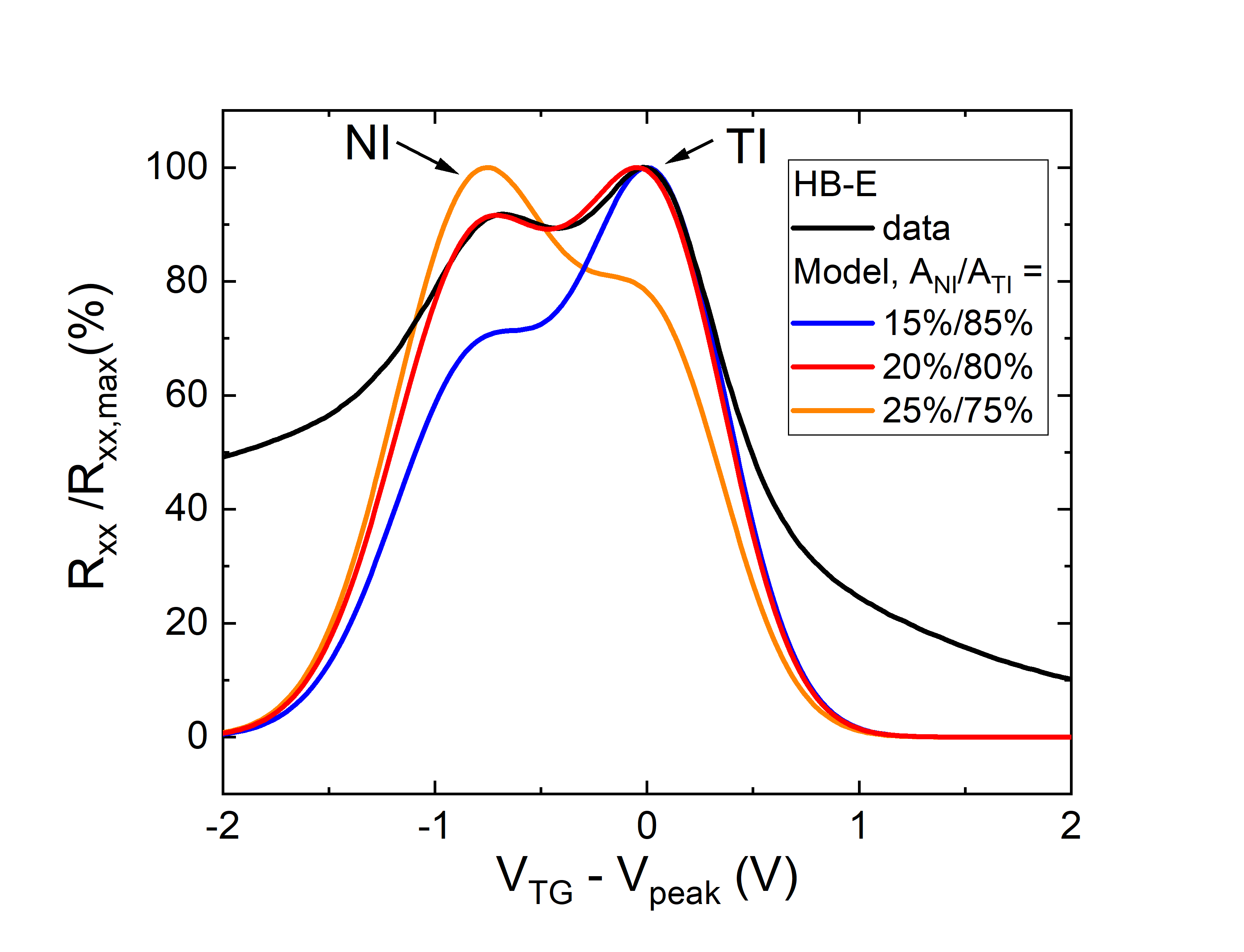} 
\caption{\label{Fig:S4} Experimental data (in black) and modeling of $R_{xx}$ for the HB-E device. All curves are normalized to the top-gate voltage and resistance value for the higher peak. The best fit is achieved with an area ratio of $A_{NI}/A_{TI}=20\%/80\%$ (in red). By increasing $A_{NI}$ to 25\% (orange) or decreasing it to 15\% (blue) the fit differs from the experimental data.}
\end{figure}

Furthermore, $V_{NI}$ and $V_{TI}$ are the voltage positions of the band-gap and $S_{NI}$ and $S_{TI}$ describe the width of the peaks. For modeling the actual double peak shape, we chose $\rho_{max,TI}=4$~k$\Omega$ (extracted from the measurements for HB-O device). As $E_{gap}(NI){\approx}4E_{gap}(TI)$ (see the main text), we determine $\rho_{NI,max}{\approx}4\rho_{TI,max}=16$~k$\Omega$. Estimating the difference in $V_{TI}$ and $V_{NI}$ as around 0.8~V (see Fig~1(c) from the main text), one may conclude that $V_{TI}=0$~V and $V_{NI}=-0.8$~V. The other remaining values can be obtained by a double Gaussian fit for both peaks of the HB-E device, which gives $S_{NI}=0.95$ and $S_{TI}=0.42$. With these parameters, the double peak for HB-E device can be qualitatively modeled by varying the ratio of the area coverage of the NI and TI phase.

The data for the HB-E device (taken from Fig 1(c) in the main text, in black) together with a few selected curves for different area ratios from the resistor network model is presented in Fig.~\ref{Fig:S4}. All curves are normalized to the top-gate voltage and resistance value for the TI peak. For a suitable fit of the model with the data, roughly 80\% of the area of the Hall bar needs to be covered by a TI phase whereas 20\% by a NI phase (red). Decreasing the area for the NI phase to $A_{NI}=15$\% (blue) or increasing it to $A_{NI}=25$\% (orange), leads to a significant deterioration of the agreement between the model and the experiment. Also, in order to observe a double-peak structure, the TI and NI phases need to be put in a series connection randomly distributed over the complete width of the Hall bar. Otherwise, only one peak is prominent, and the second peak is visible at most as a shoulder. Therefore, the resistor model allows us to ascertain the ratio of the area coverage between both phases for HB-E device and to get an understanding of how both phases need to be arranged to observe the double peak. It indicates that only the ratio between the areas and the series connection is important, while the exact arrangement has no significance.


\begin{thebibliography}{40}%
\makeatletter
\providecommand \@ifxundefined [1]{%
 \@ifx{#1\undefined}
}%
\providecommand \@ifnum [1]{%
 \ifnum #1\expandafter \@firstoftwo
 \else \expandafter \@secondoftwo
 \fi
}%
\providecommand \@ifx [1]{%
 \ifx #1\expandafter \@firstoftwo
 \else \expandafter \@secondoftwo
 \fi
}%
\providecommand \natexlab [1]{#1}%
\providecommand \enquote  [1]{``#1''}%
\providecommand \bibnamefont  [1]{#1}%
\providecommand \bibfnamefont [1]{#1}%
\providecommand \citenamefont [1]{#1}%
\providecommand \href@noop [0]{\@secondoftwo}%
\providecommand \href [0]{\begingroup \@sanitize@url \@href}%
\providecommand \@href[1]{\@@startlink{#1}\@@href}%
\providecommand \@@href[1]{\endgroup#1\@@endlink}%
\providecommand \@sanitize@url [0]{\catcode `\\12\catcode `\$12\catcode
  `\&12\catcode `\#12\catcode `\^12\catcode `\_12\catcode `\%12\relax}%
\providecommand \@@startlink[1]{}%
\providecommand \@@endlink[0]{}%
\providecommand \url  [0]{\begingroup\@sanitize@url \@url }%
\providecommand \@url [1]{\endgroup\@href {#1}{\urlprefix }}%
\providecommand \urlprefix  [0]{URL }%
\providecommand \Eprint [0]{\href }%
\providecommand \doibase [0]{http://dx.doi.org/}%
\providecommand \selectlanguage [0]{\@gobble}%
\providecommand \bibinfo  [0]{\@secondoftwo}%
\providecommand \bibfield  [0]{\@secondoftwo}%
\providecommand \translation [1]{[#1]}%
\providecommand \BibitemOpen [0]{}%
\providecommand \bibitemStop [0]{}%
\providecommand \bibitemNoStop [0]{.\EOS\space}%
\providecommand \EOS [0]{\spacefactor3000\relax}%
\providecommand \BibitemShut  [1]{\csname bibitem#1\endcsname}%
\let\auto@bib@innerbib\@empty
\bibitem [{\citenamefont {Kane}\ and\ \citenamefont {Mele}(2005)}]{ref1}%
  \BibitemOpen
  \bibfield  {author} {\bibinfo {author} {\bibfnamefont {C.~L.}\ \bibnamefont
  {Kane}}\ and\ \bibinfo {author} {\bibfnamefont {E.~J.}\ \bibnamefont
  {Mele}},\ }\href {\doibase 10.1103/PhysRevLett.95.146802} {\bibfield
  {journal} {\bibinfo  {journal} {Phys. Rev. Lett.}\ }\textbf {\bibinfo
  {volume} {95}},\ \bibinfo {pages} {146802} (\bibinfo {year}
  {2005})}\BibitemShut {NoStop}%
\bibitem [{\citenamefont {Bernevig}\ \emph {et~al.}(2006)\citenamefont
  {Bernevig}, \citenamefont {Hughes},\ and\ \citenamefont {Zhang}}]{ref2}%
  \BibitemOpen
  \bibfield  {author} {\bibinfo {author} {\bibfnamefont {B.~A.}\ \bibnamefont
  {Bernevig}}, \bibinfo {author} {\bibfnamefont {T.~L.}\ \bibnamefont
  {Hughes}}, \ and\ \bibinfo {author} {\bibfnamefont {S.-C.}\ \bibnamefont
  {Zhang}},\ }\href {\doibase 10.1126/science.1133734} {\bibfield  {journal}
  {\bibinfo  {journal} {Science}\ }\textbf {\bibinfo {volume} {314}},\ \bibinfo
  {pages} {1757} (\bibinfo {year} {2006})}\BibitemShut {NoStop}%
\bibitem [{\citenamefont {K\"{o}nig}\ \emph {et~al.}(2007)\citenamefont
  {K\"{o}nig}, \citenamefont {Wiedmann}, \citenamefont {Br\"{u}ne},
  \citenamefont {Roth}, \citenamefont {Buhmann}, \citenamefont {Molenkamp},
  \citenamefont {Qi},\ and\ \citenamefont {Zhang}}]{ref3}%
  \BibitemOpen
  \bibfield  {author} {\bibinfo {author} {\bibfnamefont {M.}~\bibnamefont
  {K\"{o}nig}}, \bibinfo {author} {\bibfnamefont {S.}~\bibnamefont {Wiedmann}},
  \bibinfo {author} {\bibfnamefont {C.}~\bibnamefont {Br\"{u}ne}}, \bibinfo
  {author} {\bibfnamefont {A.}~\bibnamefont {Roth}}, \bibinfo {author}
  {\bibfnamefont {H.}~\bibnamefont {Buhmann}}, \bibinfo {author} {\bibfnamefont
  {L.~W.}\ \bibnamefont {Molenkamp}}, \bibinfo {author} {\bibfnamefont {X.-L.}\
  \bibnamefont {Qi}}, \ and\ \bibinfo {author} {\bibfnamefont {S.-C.}\
  \bibnamefont {Zhang}},\ }\href {\doibase 10.1126/science.1148047} {\bibfield
  {journal} {\bibinfo  {journal} {Science}\ }\textbf {\bibinfo {volume}
  {318}},\ \bibinfo {pages} {766} (\bibinfo {year} {2007})}\BibitemShut
  {NoStop}%
\bibitem [{\citenamefont {Roth}\ \emph {et~al.}(2009)\citenamefont {Roth},
  \citenamefont {Br\"{u}ne}, \citenamefont {Buhmann}, \citenamefont
  {Molenkamp}, \citenamefont {Maciejko}, \citenamefont {Qi},\ and\
  \citenamefont {Zhang}}]{ref4}%
  \BibitemOpen
  \bibfield  {author} {\bibinfo {author} {\bibfnamefont {A.}~\bibnamefont
  {Roth}}, \bibinfo {author} {\bibfnamefont {C.}~\bibnamefont {Br\"{u}ne}},
  \bibinfo {author} {\bibfnamefont {H.}~\bibnamefont {Buhmann}}, \bibinfo
  {author} {\bibfnamefont {L.~W.}\ \bibnamefont {Molenkamp}}, \bibinfo {author}
  {\bibfnamefont {J.}~\bibnamefont {Maciejko}}, \bibinfo {author}
  {\bibfnamefont {X.-L.}\ \bibnamefont {Qi}}, \ and\ \bibinfo {author}
  {\bibfnamefont {S.-C.}\ \bibnamefont {Zhang}},\ }\href {\doibase
  10.1126/science.1174736} {\bibfield  {journal} {\bibinfo  {journal}
  {Science}\ }\textbf {\bibinfo {volume} {325}},\ \bibinfo {pages} {294}
  (\bibinfo {year} {2009})}\BibitemShut {NoStop}%
\bibitem [{\citenamefont {Hasan}\ and\ \citenamefont {Kane}(2010)}]{ref5}%
  \BibitemOpen
  \bibfield  {author} {\bibinfo {author} {\bibfnamefont {M.~Z.}\ \bibnamefont
  {Hasan}}\ and\ \bibinfo {author} {\bibfnamefont {C.~L.}\ \bibnamefont
  {Kane}},\ }\href {\doibase 10.1103/RevModPhys.82.3045} {\bibfield  {journal}
  {\bibinfo  {journal} {Rev. Mod. Phys.}\ }\textbf {\bibinfo {volume} {82}},\
  \bibinfo {pages} {3045} (\bibinfo {year} {2010})}\BibitemShut {NoStop}%
\bibitem [{\citenamefont {Wu}\ \emph {et~al.}(2018)\citenamefont {Wu},
  \citenamefont {Fatemi}, \citenamefont {Gibson}, \citenamefont {Watanabe},
  \citenamefont {Taniguchi}, \citenamefont {Cava},\ and\ \citenamefont
  {Jarillo-Herrero}}]{ref6}%
  \BibitemOpen
  \bibfield  {author} {\bibinfo {author} {\bibfnamefont {S.}~\bibnamefont
  {Wu}}, \bibinfo {author} {\bibfnamefont {V.}~\bibnamefont {Fatemi}}, \bibinfo
  {author} {\bibfnamefont {Q.~D.}\ \bibnamefont {Gibson}}, \bibinfo {author}
  {\bibfnamefont {K.}~\bibnamefont {Watanabe}}, \bibinfo {author}
  {\bibfnamefont {T.}~\bibnamefont {Taniguchi}}, \bibinfo {author}
  {\bibfnamefont {R.~J.}\ \bibnamefont {Cava}}, \ and\ \bibinfo {author}
  {\bibfnamefont {P.}~\bibnamefont {Jarillo-Herrero}},\ }\href {\doibase
  10.1126/science.aan6003} {\bibfield  {journal} {\bibinfo  {journal}
  {Science}\ }\textbf {\bibinfo {volume} {359}},\ \bibinfo {pages} {76}
  (\bibinfo {year} {2018})}\BibitemShut {NoStop}%
\bibitem [{\citenamefont {Zhang}\ \emph {et~al.}(2009)\citenamefont {Zhang},
  \citenamefont {Liu}, \citenamefont {Qi}, \citenamefont {Dai}, \citenamefont
  {Fang},\ and\ \citenamefont {Zhang}}]{ref7}%
  \BibitemOpen
  \bibfield  {author} {\bibinfo {author} {\bibfnamefont {H.}~\bibnamefont
  {Zhang}}, \bibinfo {author} {\bibfnamefont {C.-X.}\ \bibnamefont {Liu}},
  \bibinfo {author} {\bibfnamefont {X.-L.}\ \bibnamefont {Qi}}, \bibinfo
  {author} {\bibfnamefont {X.}~\bibnamefont {Dai}}, \bibinfo {author}
  {\bibfnamefont {Z.}~\bibnamefont {Fang}}, \ and\ \bibinfo {author}
  {\bibfnamefont {S.-C.}\ \bibnamefont {Zhang}},\ }\href {\doibase
  10.1038/nphys1270} {\bibfield  {journal} {\bibinfo  {journal} {Nat. Phys.}\
  }\textbf {\bibinfo {volume} {5}},\ \bibinfo {pages} {438} (\bibinfo {year}
  {2009})}\BibitemShut {NoStop}%
\bibitem [{\citenamefont {Vergniory}\ \emph {et~al.}(2019)\citenamefont
  {Vergniory}, \citenamefont {Elcoro}, \citenamefont {Felser}, \citenamefont
  {Regnault}, \citenamefont {Bernevig},\ and\ \citenamefont {Wang}}]{ref8}%
  \BibitemOpen
  \bibfield  {author} {\bibinfo {author} {\bibfnamefont {M.~G.}\ \bibnamefont
  {Vergniory}}, \bibinfo {author} {\bibfnamefont {L.}~\bibnamefont {Elcoro}},
  \bibinfo {author} {\bibfnamefont {C.}~\bibnamefont {Felser}}, \bibinfo
  {author} {\bibfnamefont {N.}~\bibnamefont {Regnault}}, \bibinfo {author}
  {\bibfnamefont {B.~A.}\ \bibnamefont {Bernevig}}, \ and\ \bibinfo {author}
  {\bibfnamefont {Z.}~\bibnamefont {Wang}},\ }\href {\doibase
  10.1038/s41586-019-0954-4} {\bibfield  {journal} {\bibinfo  {journal}
  {Nature}\ }\textbf {\bibinfo {volume} {566}},\ \bibinfo {pages} {480}
  (\bibinfo {year} {2019})}\BibitemShut {NoStop}%
\bibitem [{\citenamefont {Liu}\ \emph {et~al.}(2008)\citenamefont {Liu},
  \citenamefont {Hughes}, \citenamefont {Qi}, \citenamefont {Wang},\ and\
  \citenamefont {Zhang}}]{ref9}%
  \BibitemOpen
  \bibfield  {author} {\bibinfo {author} {\bibfnamefont {C.}~\bibnamefont
  {Liu}}, \bibinfo {author} {\bibfnamefont {T.~L.}\ \bibnamefont {Hughes}},
  \bibinfo {author} {\bibfnamefont {X.-L.}\ \bibnamefont {Qi}}, \bibinfo
  {author} {\bibfnamefont {K.}~\bibnamefont {Wang}}, \ and\ \bibinfo {author}
  {\bibfnamefont {S.-C.}\ \bibnamefont {Zhang}},\ }\href {\doibase
  10.1103/PhysRevLett.100.236601} {\bibfield  {journal} {\bibinfo  {journal}
  {Phys. Rev. Lett.}\ }\textbf {\bibinfo {volume} {100}},\ \bibinfo {pages}
  {236601} (\bibinfo {year} {2008})}\BibitemShut {NoStop}%
\bibitem [{\citenamefont {Tourni\'{e}}\ \emph {et~al.}(2022)\citenamefont
  {Tourni\'{e}}, \citenamefont {Bartolome}, \citenamefont {Calvo},
  \citenamefont {Loghmari}, \citenamefont {D\'{i}az-Thomas}, \citenamefont
  {Teissier}, \citenamefont {Baranov}, \citenamefont {Cerutti},\ and\
  \citenamefont {Rodriguez}}]{ref10}%
  \BibitemOpen
  \bibfield  {author} {\bibinfo {author} {\bibfnamefont {E.}~\bibnamefont
  {Tourni\'{e}}}, \bibinfo {author} {\bibfnamefont {L.~M.}\ \bibnamefont
  {Bartolome}}, \bibinfo {author} {\bibfnamefont {M.~R.}\ \bibnamefont
  {Calvo}}, \bibinfo {author} {\bibfnamefont {Z.}~\bibnamefont {Loghmari}},
  \bibinfo {author} {\bibfnamefont {D.~A.}\ \bibnamefont {D\'{i}az-Thomas}},
  \bibinfo {author} {\bibfnamefont {R.}~\bibnamefont {Teissier}}, \bibinfo
  {author} {\bibfnamefont {A.~N.}\ \bibnamefont {Baranov}}, \bibinfo {author}
  {\bibfnamefont {L.}~\bibnamefont {Cerutti}}, \ and\ \bibinfo {author}
  {\bibfnamefont {J.-B.}\ \bibnamefont {Rodriguez}},\ }\href {\doibase
  10.1038/s41377-022-00850-4} {\bibfield  {journal} {\bibinfo  {journal} {Light
  Sci. Appl.}\ }\textbf {\bibinfo {volume} {11}},\ \bibinfo {pages} {165}
  (\bibinfo {year} {2022})}\BibitemShut {NoStop}%
\bibitem [{\citenamefont {Knez}\ \emph {et~al.}(2011)\citenamefont {Knez},
  \citenamefont {Du},\ and\ \citenamefont {Sullivan}}]{ref11}%
  \BibitemOpen
  \bibfield  {author} {\bibinfo {author} {\bibfnamefont {I.}~\bibnamefont
  {Knez}}, \bibinfo {author} {\bibfnamefont {R.-R.}\ \bibnamefont {Du}}, \ and\
  \bibinfo {author} {\bibfnamefont {G.}~\bibnamefont {Sullivan}},\ }\href
  {\doibase 10.1103/PhysRevLett.107.136603} {\bibfield  {journal} {\bibinfo
  {journal} {Phys. Rev. Lett.}\ }\textbf {\bibinfo {volume} {107}},\ \bibinfo
  {pages} {136603} (\bibinfo {year} {2011})}\BibitemShut {NoStop}%
\bibitem [{\citenamefont {Du}\ \emph {et~al.}(2015)\citenamefont {Du},
  \citenamefont {Knez}, \citenamefont {Sullivan},\ and\ \citenamefont
  {Du}}]{ref12}%
  \BibitemOpen
  \bibfield  {author} {\bibinfo {author} {\bibfnamefont {L.}~\bibnamefont
  {Du}}, \bibinfo {author} {\bibfnamefont {I.}~\bibnamefont {Knez}}, \bibinfo
  {author} {\bibfnamefont {G.}~\bibnamefont {Sullivan}}, \ and\ \bibinfo
  {author} {\bibfnamefont {R.-R.}\ \bibnamefont {Du}},\ }\href {\doibase
  10.1103/PhysRevLett.114.096802} {\bibfield  {journal} {\bibinfo  {journal}
  {Phys. Rev. Lett.}\ }\textbf {\bibinfo {volume} {114}},\ \bibinfo {pages}
  {096802} (\bibinfo {year} {2015})}\BibitemShut {NoStop}%
\bibitem [{\citenamefont {Qu}\ \emph {et~al.}(2015)\citenamefont {Qu},
  \citenamefont {Beukman}, \citenamefont {Nadj-Perge}, \citenamefont {Wimmer},
  \citenamefont {Nguyen}, \citenamefont {Yi}, \citenamefont {Thorp},
  \citenamefont {Sokolich}, \citenamefont {Kiselev}, \citenamefont {Manfra},
  \citenamefont {Marcus},\ and\ \citenamefont {Kouwenhoven}}]{ref13}%
  \BibitemOpen
  \bibfield  {author} {\bibinfo {author} {\bibfnamefont {F.}~\bibnamefont
  {Qu}}, \bibinfo {author} {\bibfnamefont {A.~J.~A.}\ \bibnamefont {Beukman}},
  \bibinfo {author} {\bibfnamefont {S.}~\bibnamefont {Nadj-Perge}}, \bibinfo
  {author} {\bibfnamefont {M.}~\bibnamefont {Wimmer}}, \bibinfo {author}
  {\bibfnamefont {B.-M.}\ \bibnamefont {Nguyen}}, \bibinfo {author}
  {\bibfnamefont {W.}~\bibnamefont {Yi}}, \bibinfo {author} {\bibfnamefont
  {J.}~\bibnamefont {Thorp}}, \bibinfo {author} {\bibfnamefont
  {M.}~\bibnamefont {Sokolich}}, \bibinfo {author} {\bibfnamefont {A.~A.}\
  \bibnamefont {Kiselev}}, \bibinfo {author} {\bibfnamefont {M.~J.}\
  \bibnamefont {Manfra}}, \bibinfo {author} {\bibfnamefont {C.~M.}\
  \bibnamefont {Marcus}}, \ and\ \bibinfo {author} {\bibfnamefont {L.~P.}\
  \bibnamefont {Kouwenhoven}},\ }\href {\doibase
  10.1103/PhysRevLett.115.036803} {\bibfield  {journal} {\bibinfo  {journal}
  {Phys. Rev. Lett.}\ }\textbf {\bibinfo {volume} {115}},\ \bibinfo {pages}
  {036803} (\bibinfo {year} {2015})}\BibitemShut {NoStop}%
\bibitem [{\citenamefont {Dietl}(2023)}]{ref14}%
  \BibitemOpen
  \bibfield  {author} {\bibinfo {author} {\bibfnamefont {T.}~\bibnamefont
  {Dietl}},\ }\href {\doibase 10.1103/PhysRevLett.130.086202} {\bibfield
  {journal} {\bibinfo  {journal} {Phys. Rev. Lett.}\ }\textbf {\bibinfo
  {volume} {130}},\ \bibinfo {pages} {086202} (\bibinfo {year}
  {2023})}\BibitemShut {NoStop}%
\bibitem [{\citenamefont {Avogadri}\ \emph {et~al.}(2022)\citenamefont
  {Avogadri}, \citenamefont {Gebert}, \citenamefont {Krishtopenko},
  \citenamefont {Castillo}, \citenamefont {Consejo}, \citenamefont {Ruffenach},
  \citenamefont {Roblin}, \citenamefont {Bray}, \citenamefont {Krupko},
  \citenamefont {Juillaguet}, \citenamefont {Contreras}, \citenamefont {Wolf},
  \citenamefont {Hartmann}, \citenamefont {H\"ofling}, \citenamefont
  {Boissier}, \citenamefont {Rodriguez}, \citenamefont {Nanot}, \citenamefont
  {Tourni\'e}, \citenamefont {Teppe},\ and\ \citenamefont {Jouault}}]{ref15}%
  \BibitemOpen
  \bibfield  {author} {\bibinfo {author} {\bibfnamefont {C.}~\bibnamefont
  {Avogadri}}, \bibinfo {author} {\bibfnamefont {S.}~\bibnamefont {Gebert}},
  \bibinfo {author} {\bibfnamefont {S.~S.}\ \bibnamefont {Krishtopenko}},
  \bibinfo {author} {\bibfnamefont {I.}~\bibnamefont {Castillo}}, \bibinfo
  {author} {\bibfnamefont {C.}~\bibnamefont {Consejo}}, \bibinfo {author}
  {\bibfnamefont {S.}~\bibnamefont {Ruffenach}}, \bibinfo {author}
  {\bibfnamefont {C.}~\bibnamefont {Roblin}}, \bibinfo {author} {\bibfnamefont
  {C.}~\bibnamefont {Bray}}, \bibinfo {author} {\bibfnamefont {Y.}~\bibnamefont
  {Krupko}}, \bibinfo {author} {\bibfnamefont {S.}~\bibnamefont {Juillaguet}},
  \bibinfo {author} {\bibfnamefont {S.}~\bibnamefont {Contreras}}, \bibinfo
  {author} {\bibfnamefont {A.}~\bibnamefont {Wolf}}, \bibinfo {author}
  {\bibfnamefont {F.}~\bibnamefont {Hartmann}}, \bibinfo {author}
  {\bibfnamefont {S.}~\bibnamefont {H\"ofling}}, \bibinfo {author}
  {\bibfnamefont {G.}~\bibnamefont {Boissier}}, \bibinfo {author}
  {\bibfnamefont {J.-B.}\ \bibnamefont {Rodriguez}}, \bibinfo {author}
  {\bibfnamefont {S.}~\bibnamefont {Nanot}}, \bibinfo {author} {\bibfnamefont
  {E.}~\bibnamefont {Tourni\'e}}, \bibinfo {author} {\bibfnamefont
  {F.}~\bibnamefont {Teppe}}, \ and\ \bibinfo {author} {\bibfnamefont
  {B.}~\bibnamefont {Jouault}},\ }\href {\doibase
  10.1103/PhysRevResearch.4.L042042} {\bibfield  {journal} {\bibinfo  {journal}
  {Phys. Rev. Res.}\ }\textbf {\bibinfo {volume} {4}},\ \bibinfo {pages}
  {L042042} (\bibinfo {year} {2022})}\BibitemShut {NoStop}%
\bibitem [{\citenamefont {Knez}\ \emph {et~al.}(2010)\citenamefont {Knez},
  \citenamefont {Du},\ and\ \citenamefont {Sullivan}}]{ref16}%
  \BibitemOpen
  \bibfield  {author} {\bibinfo {author} {\bibfnamefont {I.}~\bibnamefont
  {Knez}}, \bibinfo {author} {\bibfnamefont {R.~R.}\ \bibnamefont {Du}}, \ and\
  \bibinfo {author} {\bibfnamefont {G.}~\bibnamefont {Sullivan}},\ }\href
  {\doibase 10.1103/PhysRevB.81.201301} {\bibfield  {journal} {\bibinfo
  {journal} {Phys. Rev. B}\ }\textbf {\bibinfo {volume} {81}},\ \bibinfo
  {pages} {201301} (\bibinfo {year} {2010})}\BibitemShut {NoStop}%
\bibitem [{\citenamefont {Nichele}\ \emph {et~al.}(2016)\citenamefont
  {Nichele}, \citenamefont {Suominen}, \citenamefont {Kjaergaard},
  \citenamefont {Marcus}, \citenamefont {Sajadi}, \citenamefont {Folk},
  \citenamefont {Qu}, \citenamefont {Beukman}, \citenamefont {de~Vries},
  \citenamefont {van Veen}, \citenamefont {Nadj-Perge}, \citenamefont
  {Kouwenhoven}, \citenamefont {Nguyen}, \citenamefont {Kiselev}, \citenamefont
  {Yi}, \citenamefont {Sokolich}, \citenamefont {Manfra}, \citenamefont
  {Spanton},\ and\ \citenamefont {Moler}}]{ref17}%
  \BibitemOpen
  \bibfield  {author} {\bibinfo {author} {\bibfnamefont {F.}~\bibnamefont
  {Nichele}}, \bibinfo {author} {\bibfnamefont {H.~J.}\ \bibnamefont
  {Suominen}}, \bibinfo {author} {\bibfnamefont {M.}~\bibnamefont
  {Kjaergaard}}, \bibinfo {author} {\bibfnamefont {C.~M.}\ \bibnamefont
  {Marcus}}, \bibinfo {author} {\bibfnamefont {E.}~\bibnamefont {Sajadi}},
  \bibinfo {author} {\bibfnamefont {J.~A.}\ \bibnamefont {Folk}}, \bibinfo
  {author} {\bibfnamefont {F.}~\bibnamefont {Qu}}, \bibinfo {author}
  {\bibfnamefont {A.~J.~A.}\ \bibnamefont {Beukman}}, \bibinfo {author}
  {\bibfnamefont {F.~K.}\ \bibnamefont {de~Vries}}, \bibinfo {author}
  {\bibfnamefont {J.}~\bibnamefont {van Veen}}, \bibinfo {author}
  {\bibfnamefont {S.}~\bibnamefont {Nadj-Perge}}, \bibinfo {author}
  {\bibfnamefont {L.~P.}\ \bibnamefont {Kouwenhoven}}, \bibinfo {author}
  {\bibfnamefont {B.-M.}\ \bibnamefont {Nguyen}}, \bibinfo {author}
  {\bibfnamefont {A.~A.}\ \bibnamefont {Kiselev}}, \bibinfo {author}
  {\bibfnamefont {W.}~\bibnamefont {Yi}}, \bibinfo {author} {\bibfnamefont
  {M.}~\bibnamefont {Sokolich}}, \bibinfo {author} {\bibfnamefont {M.~J.}\
  \bibnamefont {Manfra}}, \bibinfo {author} {\bibfnamefont {E.~M.}\
  \bibnamefont {Spanton}}, \ and\ \bibinfo {author} {\bibfnamefont {K.~A.}\
  \bibnamefont {Moler}},\ }\href {\doibase 10.1088/1367-2630/18/8/083005}
  {\bibfield  {journal} {\bibinfo  {journal} {New J. Phys.}\ }\textbf {\bibinfo
  {volume} {18}},\ \bibinfo {pages} {083005} (\bibinfo {year}
  {2016})}\BibitemShut {NoStop}%
\bibitem [{\citenamefont {Mittag}\ \emph {et~al.}(2017)\citenamefont {Mittag},
  \citenamefont {Karalic}, \citenamefont {Mueller}, \citenamefont {Tschirky},
  \citenamefont {Wegscheider}, \citenamefont {Nazarenko}, \citenamefont
  {Kovalenko}, \citenamefont {Ihn},\ and\ \citenamefont {Ensslin}}]{ref18}%
  \BibitemOpen
  \bibfield  {author} {\bibinfo {author} {\bibfnamefont {C.}~\bibnamefont
  {Mittag}}, \bibinfo {author} {\bibfnamefont {M.}~\bibnamefont {Karalic}},
  \bibinfo {author} {\bibfnamefont {S.}~\bibnamefont {Mueller}}, \bibinfo
  {author} {\bibfnamefont {T.}~\bibnamefont {Tschirky}}, \bibinfo {author}
  {\bibfnamefont {W.}~\bibnamefont {Wegscheider}}, \bibinfo {author}
  {\bibfnamefont {O.}~\bibnamefont {Nazarenko}}, \bibinfo {author}
  {\bibfnamefont {M.~V.}\ \bibnamefont {Kovalenko}}, \bibinfo {author}
  {\bibfnamefont {T.}~\bibnamefont {Ihn}}, \ and\ \bibinfo {author}
  {\bibfnamefont {K.}~\bibnamefont {Ensslin}},\ }\href {\doibase
  10.1063/1.4986614} {\bibfield  {journal} {\bibinfo  {journal} {Appl. Phys.
  Lett.}\ }\textbf {\bibinfo {volume} {111}},\ \bibinfo {pages} {082101}
  (\bibinfo {year} {2017})}\BibitemShut {NoStop}%
\bibitem [{\citenamefont {Mueller}\ \emph {et~al.}(2017)\citenamefont
  {Mueller}, \citenamefont {Mittag}, \citenamefont {Tschirky}, \citenamefont
  {Charpentier}, \citenamefont {Wegscheider}, \citenamefont {Ensslin},\ and\
  \citenamefont {Ihn}}]{ref19}%
  \BibitemOpen
  \bibfield  {author} {\bibinfo {author} {\bibfnamefont {S.}~\bibnamefont
  {Mueller}}, \bibinfo {author} {\bibfnamefont {C.}~\bibnamefont {Mittag}},
  \bibinfo {author} {\bibfnamefont {T.}~\bibnamefont {Tschirky}}, \bibinfo
  {author} {\bibfnamefont {C.}~\bibnamefont {Charpentier}}, \bibinfo {author}
  {\bibfnamefont {W.}~\bibnamefont {Wegscheider}}, \bibinfo {author}
  {\bibfnamefont {K.}~\bibnamefont {Ensslin}}, \ and\ \bibinfo {author}
  {\bibfnamefont {T.}~\bibnamefont {Ihn}},\ }\href {\doibase
  10.1103/PhysRevB.96.075406} {\bibfield  {journal} {\bibinfo  {journal} {Phys.
  Rev. B}\ }\textbf {\bibinfo {volume} {96}},\ \bibinfo {pages} {075406}
  (\bibinfo {year} {2017})}\BibitemShut {NoStop}%
\bibitem [{\citenamefont {Mueller}\ \emph {et~al.}(2015)\citenamefont
  {Mueller}, \citenamefont {Pal}, \citenamefont {Karalic}, \citenamefont
  {Tschirky}, \citenamefont {Charpentier}, \citenamefont {Wegscheider},
  \citenamefont {Ensslin},\ and\ \citenamefont {Ihn}}]{ref20}%
  \BibitemOpen
  \bibfield  {author} {\bibinfo {author} {\bibfnamefont {S.}~\bibnamefont
  {Mueller}}, \bibinfo {author} {\bibfnamefont {A.~N.}\ \bibnamefont {Pal}},
  \bibinfo {author} {\bibfnamefont {M.}~\bibnamefont {Karalic}}, \bibinfo
  {author} {\bibfnamefont {T.}~\bibnamefont {Tschirky}}, \bibinfo {author}
  {\bibfnamefont {C.}~\bibnamefont {Charpentier}}, \bibinfo {author}
  {\bibfnamefont {W.}~\bibnamefont {Wegscheider}}, \bibinfo {author}
  {\bibfnamefont {K.}~\bibnamefont {Ensslin}}, \ and\ \bibinfo {author}
  {\bibfnamefont {T.}~\bibnamefont {Ihn}},\ }\href {\doibase
  10.1103/PhysRevB.92.081303} {\bibfield  {journal} {\bibinfo  {journal} {Phys.
  Rev. B}\ }\textbf {\bibinfo {volume} {92}},\ \bibinfo {pages} {081303}
  (\bibinfo {year} {2015})}\BibitemShut {NoStop}%
\bibitem [{\citenamefont {Knebl}\ \emph {et~al.}(2018)\citenamefont {Knebl},
  \citenamefont {Pfeffer}, \citenamefont {Schmid}, \citenamefont {Kamp},
  \citenamefont {Bastard}, \citenamefont {Batke}, \citenamefont {Worschech},
  \citenamefont {Hartmann},\ and\ \citenamefont {H\"ofling}}]{ref21}%
  \BibitemOpen
  \bibfield  {author} {\bibinfo {author} {\bibfnamefont {G.}~\bibnamefont
  {Knebl}}, \bibinfo {author} {\bibfnamefont {P.}~\bibnamefont {Pfeffer}},
  \bibinfo {author} {\bibfnamefont {S.}~\bibnamefont {Schmid}}, \bibinfo
  {author} {\bibfnamefont {M.}~\bibnamefont {Kamp}}, \bibinfo {author}
  {\bibfnamefont {G.}~\bibnamefont {Bastard}}, \bibinfo {author} {\bibfnamefont
  {E.}~\bibnamefont {Batke}}, \bibinfo {author} {\bibfnamefont
  {L.}~\bibnamefont {Worschech}}, \bibinfo {author} {\bibfnamefont
  {F.}~\bibnamefont {Hartmann}}, \ and\ \bibinfo {author} {\bibfnamefont
  {S.}~\bibnamefont {H\"ofling}},\ }\href {\doibase 10.1103/PhysRevB.98.041301}
  {\bibfield  {journal} {\bibinfo  {journal} {Phys. Rev. B}\ }\textbf {\bibinfo
  {volume} {98}},\ \bibinfo {pages} {041301} (\bibinfo {year}
  {2018})}\BibitemShut {NoStop}%
\bibitem [{\citenamefont {Schets}\ \emph {et~al.}(1999)\citenamefont {Schets},
  \citenamefont {Janssen}, \citenamefont {Witters},\ and\ \citenamefont
  {Borghs}}]{ref22}%
  \BibitemOpen
  \bibfield  {author} {\bibinfo {author} {\bibfnamefont {H.}~\bibnamefont
  {Schets}}, \bibinfo {author} {\bibfnamefont {P.}~\bibnamefont {Janssen}},
  \bibinfo {author} {\bibfnamefont {J.}~\bibnamefont {Witters}}, \ and\
  \bibinfo {author} {\bibfnamefont {S.}~\bibnamefont {Borghs}},\ }\href
  {\doibase 10.1016/S0038-1098(98)00625-5} {\bibfield  {journal} {\bibinfo
  {journal} {Solid State Commun.}\ }\textbf {\bibinfo {volume} {110}},\
  \bibinfo {pages} {169} (\bibinfo {year} {1999})}\BibitemShut {NoStop}%
\bibitem [{ref()}]{ref23}%
  \BibitemOpen
  \href@noop {} {\bibinfo  {journal} {See Supplemental Material containing
  Refs.~[39,40] for the details of sample growth and processing, as well as the
  details of band structure calculations. Analysis of double $\rho_{xx}$ peak
  for HB-E device within a resistor network model also provided therein}\
  }\BibitemShut {NoStop}%
\bibitem [{\citenamefont {Karalic}\ \emph {et~al.}(2016)\citenamefont
  {Karalic}, \citenamefont {Mueller}, \citenamefont {Mittag}, \citenamefont
  {Pakrouski}, \citenamefont {Wu}, \citenamefont {Soluyanov}, \citenamefont
  {Troyer}, \citenamefont {Tschirky}, \citenamefont {Wegscheider},
  \citenamefont {Ensslin},\ and\ \citenamefont {Ihn}}]{ref24}%
  \BibitemOpen
\bibfield  {journal} {  }\bibfield  {author} {\bibinfo {author} {\bibfnamefont
  {M.}~\bibnamefont {Karalic}}, \bibinfo {author} {\bibfnamefont
  {S.}~\bibnamefont {Mueller}}, \bibinfo {author} {\bibfnamefont
  {C.}~\bibnamefont {Mittag}}, \bibinfo {author} {\bibfnamefont
  {K.}~\bibnamefont {Pakrouski}}, \bibinfo {author} {\bibfnamefont
  {Q.}~\bibnamefont {Wu}}, \bibinfo {author} {\bibfnamefont {A.~A.}\
  \bibnamefont {Soluyanov}}, \bibinfo {author} {\bibfnamefont {M.}~\bibnamefont
  {Troyer}}, \bibinfo {author} {\bibfnamefont {T.}~\bibnamefont {Tschirky}},
  \bibinfo {author} {\bibfnamefont {W.}~\bibnamefont {Wegscheider}}, \bibinfo
  {author} {\bibfnamefont {K.}~\bibnamefont {Ensslin}}, \ and\ \bibinfo
  {author} {\bibfnamefont {T.}~\bibnamefont {Ihn}},\ }\href {\doibase
  10.1103/PhysRevB.94.241402} {\bibfield  {journal} {\bibinfo  {journal} {Phys.
  Rev. B}\ }\textbf {\bibinfo {volume} {94}},\ \bibinfo {pages} {241402}
  (\bibinfo {year} {2016})}\BibitemShut {NoStop}%
\bibitem [{\citenamefont {Schmid}\ \emph {et~al.}(2022)\citenamefont {Schmid},
  \citenamefont {Meyer}, \citenamefont {Jabeen}, \citenamefont {Bastard},
  \citenamefont {Hartmann},\ and\ \citenamefont {H\"ofling}}]{ref25}%
  \BibitemOpen
  \bibfield  {author} {\bibinfo {author} {\bibfnamefont {S.}~\bibnamefont
  {Schmid}}, \bibinfo {author} {\bibfnamefont {M.}~\bibnamefont {Meyer}},
  \bibinfo {author} {\bibfnamefont {F.}~\bibnamefont {Jabeen}}, \bibinfo
  {author} {\bibfnamefont {G.}~\bibnamefont {Bastard}}, \bibinfo {author}
  {\bibfnamefont {F.}~\bibnamefont {Hartmann}}, \ and\ \bibinfo {author}
  {\bibfnamefont {S.}~\bibnamefont {H\"ofling}},\ }\href {\doibase
  10.1103/PhysRevB.105.155304} {\bibfield  {journal} {\bibinfo  {journal}
  {Phys. Rev. B}\ }\textbf {\bibinfo {volume} {105}},\ \bibinfo {pages}
  {155304} (\bibinfo {year} {2022})}\BibitemShut {NoStop}%
\bibitem [{\citenamefont {Tong}\ \emph {et~al.}(2017)\citenamefont {Tong},
  \citenamefont {Han}, \citenamefont {Li}, \citenamefont {Zhang}, \citenamefont
  {Sullivan},\ and\ \citenamefont {Du}}]{ref26}%
  \BibitemOpen
  \bibfield  {author} {\bibinfo {author} {\bibfnamefont {B.}~\bibnamefont
  {Tong}}, \bibinfo {author} {\bibfnamefont {Z.}~\bibnamefont {Han}}, \bibinfo
  {author} {\bibfnamefont {T.}~\bibnamefont {Li}}, \bibinfo {author}
  {\bibfnamefont {C.}~\bibnamefont {Zhang}}, \bibinfo {author} {\bibfnamefont
  {G.}~\bibnamefont {Sullivan}}, \ and\ \bibinfo {author} {\bibfnamefont
  {R.~R.}\ \bibnamefont {Du}},\ }\href {\doibase 10.1063/1.4993894} {\bibfield
  {journal} {\bibinfo  {journal} {AIP Adv.}\ }\textbf {\bibinfo {volume} {7}},\
  \bibinfo {pages} {075211} (\bibinfo {year} {2017})}\BibitemShut {NoStop}%
\bibitem [{\citenamefont {Gavrilenko}\ \emph {et~al.}(2010)\citenamefont
  {Gavrilenko}, \citenamefont {Ikonnikov}, \citenamefont {Krishtopenko},
  \citenamefont {Lastovkin}, \citenamefont {Maremyanin}, \citenamefont
  {Sadofyev},\ and\ \citenamefont {Spirin}}]{ref27}%
  \BibitemOpen
  \bibfield  {author} {\bibinfo {author} {\bibfnamefont {V.~I.}\ \bibnamefont
  {Gavrilenko}}, \bibinfo {author} {\bibfnamefont {A.~V.}\ \bibnamefont
  {Ikonnikov}}, \bibinfo {author} {\bibfnamefont {S.~S.}\ \bibnamefont
  {Krishtopenko}}, \bibinfo {author} {\bibfnamefont {A.~A.}\ \bibnamefont
  {Lastovkin}}, \bibinfo {author} {\bibfnamefont {K.~V.}\ \bibnamefont
  {Maremyanin}}, \bibinfo {author} {\bibfnamefont {Y.~G.}\ \bibnamefont
  {Sadofyev}}, \ and\ \bibinfo {author} {\bibfnamefont {K.~E.}\ \bibnamefont
  {Spirin}},\ }\href {\doibase 10.1134/S106378261005012X} {\bibfield  {journal}
  {\bibinfo  {journal} {Semiconductors}\ }\textbf {\bibinfo {volume} {44}},\
  \bibinfo {pages} {616} (\bibinfo {year} {2010})}\BibitemShut {NoStop}%
\bibitem [{\citenamefont {Gauer}\ \emph {et~al.}(1993)\citenamefont {Gauer},
  \citenamefont {Scriba}, \citenamefont {Wixforth}, \citenamefont {Kotthaus},
  \citenamefont {Nguyen}, \citenamefont {Tuttle}, \citenamefont {English},\
  and\ \citenamefont {Kroemer}}]{ref28}%
  \BibitemOpen
  \bibfield  {author} {\bibinfo {author} {\bibfnamefont {C.}~\bibnamefont
  {Gauer}}, \bibinfo {author} {\bibfnamefont {J.}~\bibnamefont {Scriba}},
  \bibinfo {author} {\bibfnamefont {A.}~\bibnamefont {Wixforth}}, \bibinfo
  {author} {\bibfnamefont {J.~P.}\ \bibnamefont {Kotthaus}}, \bibinfo {author}
  {\bibfnamefont {C.}~\bibnamefont {Nguyen}}, \bibinfo {author} {\bibfnamefont
  {G.}~\bibnamefont {Tuttle}}, \bibinfo {author} {\bibfnamefont {J.~H.}\
  \bibnamefont {English}}, \ and\ \bibinfo {author} {\bibfnamefont
  {H.}~\bibnamefont {Kroemer}},\ }\href {\doibase 10.1088/0268-1242/8/1S/031}
  {\bibfield  {journal} {\bibinfo  {journal} {Semicond. Sci. Technol.}\
  }\textbf {\bibinfo {volume} {8}},\ \bibinfo {pages} {S137} (\bibinfo {year}
  {1993})}\BibitemShut {NoStop}%
\bibitem [{\citenamefont {Krishtopenko}\ \emph {et~al.}(2019)\citenamefont
  {Krishtopenko}, \citenamefont {Desrat}, \citenamefont {Spirin}, \citenamefont
  {Consejo}, \citenamefont {Ruffenach}, \citenamefont {Gonzalez-Posada},
  \citenamefont {Jouault}, \citenamefont {Knap}, \citenamefont {Maremyanin},
  \citenamefont {Gavrilenko}, \citenamefont {Boissier}, \citenamefont {Torres},
  \citenamefont {Zaknoune}, \citenamefont {Tourni\'e},\ and\ \citenamefont
  {Teppe}}]{ref29}%
  \BibitemOpen
  \bibfield  {author} {\bibinfo {author} {\bibfnamefont {S.~S.}\ \bibnamefont
  {Krishtopenko}}, \bibinfo {author} {\bibfnamefont {W.}~\bibnamefont
  {Desrat}}, \bibinfo {author} {\bibfnamefont {K.~E.}\ \bibnamefont {Spirin}},
  \bibinfo {author} {\bibfnamefont {C.}~\bibnamefont {Consejo}}, \bibinfo
  {author} {\bibfnamefont {S.}~\bibnamefont {Ruffenach}}, \bibinfo {author}
  {\bibfnamefont {F.}~\bibnamefont {Gonzalez-Posada}}, \bibinfo {author}
  {\bibfnamefont {B.}~\bibnamefont {Jouault}}, \bibinfo {author} {\bibfnamefont
  {W.}~\bibnamefont {Knap}}, \bibinfo {author} {\bibfnamefont {K.~V.}\
  \bibnamefont {Maremyanin}}, \bibinfo {author} {\bibfnamefont {V.~I.}\
  \bibnamefont {Gavrilenko}}, \bibinfo {author} {\bibfnamefont
  {G.}~\bibnamefont {Boissier}}, \bibinfo {author} {\bibfnamefont
  {J.}~\bibnamefont {Torres}}, \bibinfo {author} {\bibfnamefont
  {M.}~\bibnamefont {Zaknoune}}, \bibinfo {author} {\bibfnamefont
  {E.}~\bibnamefont {Tourni\'e}}, \ and\ \bibinfo {author} {\bibfnamefont
  {F.}~\bibnamefont {Teppe}},\ }\href {\doibase 10.1103/PhysRevB.99.121405}
  {\bibfield  {journal} {\bibinfo  {journal} {Phys. Rev. B}\ }\textbf {\bibinfo
  {volume} {99}},\ \bibinfo {pages} {121405} (\bibinfo {year}
  {2019})}\BibitemShut {NoStop}%
\bibitem [{\citenamefont {Meyer}\ \emph {et~al.}(2021)\citenamefont {Meyer},
  \citenamefont {Schmid}, \citenamefont {Jabeen}, \citenamefont {Bastard},
  \citenamefont {Hartmann},\ and\ \citenamefont {H\"ofling}}]{ref30}%
  \BibitemOpen
  \bibfield  {author} {\bibinfo {author} {\bibfnamefont {M.}~\bibnamefont
  {Meyer}}, \bibinfo {author} {\bibfnamefont {S.}~\bibnamefont {Schmid}},
  \bibinfo {author} {\bibfnamefont {F.}~\bibnamefont {Jabeen}}, \bibinfo
  {author} {\bibfnamefont {G.}~\bibnamefont {Bastard}}, \bibinfo {author}
  {\bibfnamefont {F.}~\bibnamefont {Hartmann}}, \ and\ \bibinfo {author}
  {\bibfnamefont {S.}~\bibnamefont {H\"ofling}},\ }\href {\doibase
  10.1103/PhysRevB.104.085301} {\bibfield  {journal} {\bibinfo  {journal}
  {Phys. Rev. B}\ }\textbf {\bibinfo {volume} {104}},\ \bibinfo {pages}
  {085301} (\bibinfo {year} {2021})}\BibitemShut {NoStop}%
\bibitem [{\citenamefont {Nguyen}\ \emph {et~al.}(2015)\citenamefont {Nguyen},
  \citenamefont {Yi}, \citenamefont {Noah}, \citenamefont {Thorp},\ and\
  \citenamefont {Sokolich}}]{ref31}%
  \BibitemOpen
  \bibfield  {author} {\bibinfo {author} {\bibfnamefont {B.~M.}\ \bibnamefont
  {Nguyen}}, \bibinfo {author} {\bibfnamefont {W.}~\bibnamefont {Yi}}, \bibinfo
  {author} {\bibfnamefont {R.}~\bibnamefont {Noah}}, \bibinfo {author}
  {\bibfnamefont {J.}~\bibnamefont {Thorp}}, \ and\ \bibinfo {author}
  {\bibfnamefont {M.}~\bibnamefont {Sokolich}},\ }\href {\doibase
  10.1063/1.4906589} {\bibfield  {journal} {\bibinfo  {journal} {Appl. Phys.
  Lett.}\ }\textbf {\bibinfo {volume} {106}},\ \bibinfo {pages} {032107}
  (\bibinfo {year} {2015})}\BibitemShut {NoStop}%
\bibitem [{\citenamefont {Nguyen}\ \emph {et~al.}(2016)\citenamefont {Nguyen},
  \citenamefont {Kiselev}, \citenamefont {Noah}, \citenamefont {Yi},
  \citenamefont {Qu}, \citenamefont {Beukman}, \citenamefont {de~Vries},
  \citenamefont {van Veen}, \citenamefont {Nadj-Perge}, \citenamefont
  {Kouwenhoven}, \citenamefont {Kjaergaard}, \citenamefont {Suominen},
  \citenamefont {Nichele}, \citenamefont {Marcus}, \citenamefont {Manfra},\
  and\ \citenamefont {Sokolich}}]{ref32}%
  \BibitemOpen
  \bibfield  {author} {\bibinfo {author} {\bibfnamefont {B.-M.}\ \bibnamefont
  {Nguyen}}, \bibinfo {author} {\bibfnamefont {A.~A.}\ \bibnamefont {Kiselev}},
  \bibinfo {author} {\bibfnamefont {R.}~\bibnamefont {Noah}}, \bibinfo {author}
  {\bibfnamefont {W.}~\bibnamefont {Yi}}, \bibinfo {author} {\bibfnamefont
  {F.}~\bibnamefont {Qu}}, \bibinfo {author} {\bibfnamefont {A.~J.~A.}\
  \bibnamefont {Beukman}}, \bibinfo {author} {\bibfnamefont {F.~K.}\
  \bibnamefont {de~Vries}}, \bibinfo {author} {\bibfnamefont {J.}~\bibnamefont
  {van Veen}}, \bibinfo {author} {\bibfnamefont {S.}~\bibnamefont
  {Nadj-Perge}}, \bibinfo {author} {\bibfnamefont {L.~P.}\ \bibnamefont
  {Kouwenhoven}}, \bibinfo {author} {\bibfnamefont {M.}~\bibnamefont
  {Kjaergaard}}, \bibinfo {author} {\bibfnamefont {H.~J.}\ \bibnamefont
  {Suominen}}, \bibinfo {author} {\bibfnamefont {F.}~\bibnamefont {Nichele}},
  \bibinfo {author} {\bibfnamefont {C.~M.}\ \bibnamefont {Marcus}}, \bibinfo
  {author} {\bibfnamefont {M.~J.}\ \bibnamefont {Manfra}}, \ and\ \bibinfo
  {author} {\bibfnamefont {M.}~\bibnamefont {Sokolich}},\ }\href {\doibase
  10.1103/PhysRevLett.117.077701} {\bibfield  {journal} {\bibinfo  {journal}
  {Phys. Rev. Lett.}\ }\textbf {\bibinfo {volume} {117}},\ \bibinfo {pages}
  {077701} (\bibinfo {year} {2016})}\BibitemShut {NoStop}%
\bibitem [{\citenamefont {Yang}\ \emph {et~al.}(1997)\citenamefont {Yang},
  \citenamefont {Yang}, \citenamefont {Bennett},\ and\ \citenamefont
  {Shanabrook}}]{ref33}%
  \BibitemOpen
  \bibfield  {author} {\bibinfo {author} {\bibfnamefont {M.~J.}\ \bibnamefont
  {Yang}}, \bibinfo {author} {\bibfnamefont {C.~H.}\ \bibnamefont {Yang}},
  \bibinfo {author} {\bibfnamefont {B.~R.}\ \bibnamefont {Bennett}}, \ and\
  \bibinfo {author} {\bibfnamefont {B.~V.}\ \bibnamefont {Shanabrook}},\ }\href
  {\doibase 10.1103/PhysRevLett.78.4613} {\bibfield  {journal} {\bibinfo
  {journal} {Phys. Rev. Lett.}\ }\textbf {\bibinfo {volume} {78}},\ \bibinfo
  {pages} {4613} (\bibinfo {year} {1997})}\BibitemShut {NoStop}%
\bibitem [{\citenamefont {Choi}\ \emph {et~al.}(1988)\citenamefont {Choi},
  \citenamefont {Levine}, \citenamefont {Jarosik}, \citenamefont {Walker},\
  and\ \citenamefont {Malik}}]{ref34}%
  \BibitemOpen
  \bibfield  {author} {\bibinfo {author} {\bibfnamefont {K.~K.}\ \bibnamefont
  {Choi}}, \bibinfo {author} {\bibfnamefont {B.~F.}\ \bibnamefont {Levine}},
  \bibinfo {author} {\bibfnamefont {N.}~\bibnamefont {Jarosik}}, \bibinfo
  {author} {\bibfnamefont {J.}~\bibnamefont {Walker}}, \ and\ \bibinfo {author}
  {\bibfnamefont {R.}~\bibnamefont {Malik}},\ }\href {\doibase
  10.1103/PhysRevB.38.12362} {\bibfield  {journal} {\bibinfo  {journal} {Phys.
  Rev. B}\ }\textbf {\bibinfo {volume} {38}},\ \bibinfo {pages} {12362}
  (\bibinfo {year} {1988})}\BibitemShut {NoStop}%
\bibitem [{\citenamefont {Wu}(2017)}]{ref35a}%
  \BibitemOpen
  \bibfield  {author} {\bibinfo {author} {\bibfnamefont {X.~G.}\ \bibnamefont
  {Wu}},\ }\href {\doibase 10.1063/1.5006244} {\bibfield  {journal} {\bibinfo
  {journal} {J. Appl. Phys.}\ }\textbf {\bibinfo {volume} {122}},\ \bibinfo
  {pages} {225704} (\bibinfo {year} {2017})}\BibitemShut {NoStop}%
\bibitem [{\citenamefont {Nichele}\ \emph {et~al.}(2014)\citenamefont
  {Nichele}, \citenamefont {Pal}, \citenamefont {Pietsch}, \citenamefont {Ihn},
  \citenamefont {Ensslin}, \citenamefont {Charpentier},\ and\ \citenamefont
  {Wegscheider}}]{ref35}%
  \BibitemOpen
  \bibfield  {author} {\bibinfo {author} {\bibfnamefont {F.}~\bibnamefont
  {Nichele}}, \bibinfo {author} {\bibfnamefont {A.~N.}\ \bibnamefont {Pal}},
  \bibinfo {author} {\bibfnamefont {P.}~\bibnamefont {Pietsch}}, \bibinfo
  {author} {\bibfnamefont {T.}~\bibnamefont {Ihn}}, \bibinfo {author}
  {\bibfnamefont {K.}~\bibnamefont {Ensslin}}, \bibinfo {author} {\bibfnamefont
  {C.}~\bibnamefont {Charpentier}}, \ and\ \bibinfo {author} {\bibfnamefont
  {W.}~\bibnamefont {Wegscheider}},\ }\href {\doibase
  10.1103/PhysRevLett.112.036802} {\bibfield  {journal} {\bibinfo  {journal}
  {Phys. Rev. Lett.}\ }\textbf {\bibinfo {volume} {112}},\ \bibinfo {pages}
  {036802} (\bibinfo {year} {2014})}\BibitemShut {NoStop}%
\bibitem [{\citenamefont {Krishtopenko}\ and\ \citenamefont
  {Teppe}(2018)}]{ref36}%
  \BibitemOpen
  \bibfield  {author} {\bibinfo {author} {\bibfnamefont {S.~S.}\ \bibnamefont
  {Krishtopenko}}\ and\ \bibinfo {author} {\bibfnamefont {F.}~\bibnamefont
  {Teppe}},\ }\href {\doibase 10.1126/sciadv.aap7529} {\bibfield  {journal}
  {\bibinfo  {journal} {Sci. Adv.}\ }\textbf {\bibinfo {volume} {4}},\ \bibinfo
  {pages} {eaap7529} (\bibinfo {year} {2018})}\BibitemShut {NoStop}%
\bibitem [{\citenamefont {Krishtopenko}\ \emph {et~al.}(2012)\citenamefont
  {Krishtopenko}, \citenamefont {Kalinin}, \citenamefont {Gavrilenko},
  \citenamefont {Sadofyev},\ and\ \citenamefont {Goiran}}]{ref37}%
  \BibitemOpen
  \bibfield  {author} {\bibinfo {author} {\bibfnamefont {S.~S.}\ \bibnamefont
  {Krishtopenko}}, \bibinfo {author} {\bibfnamefont {K.~P.}\ \bibnamefont
  {Kalinin}}, \bibinfo {author} {\bibfnamefont {V.~I.}\ \bibnamefont
  {Gavrilenko}}, \bibinfo {author} {\bibfnamefont {Y.~G.}\ \bibnamefont
  {Sadofyev}}, \ and\ \bibinfo {author} {\bibfnamefont {M.}~\bibnamefont
  {Goiran}},\ }\href {\doibase 10.1134/S1063782612090138} {\bibfield  {journal}
  {\bibinfo  {journal} {Semiconductors}\ }\textbf {\bibinfo {volume} {46}},\
  \bibinfo {pages} {1163} (\bibinfo {year} {2012})}\BibitemShut {NoStop}%
\bibitem [{\citenamefont {Krishtopenko}\ \emph {et~al.}(2016)\citenamefont
  {Krishtopenko}, \citenamefont {Yahniuk}, \citenamefont {But}, \citenamefont
  {Gavrilenko}, \citenamefont {Knap},\ and\ \citenamefont {Teppe}}]{SMref1q}%
  \BibitemOpen
  \bibfield  {author} {\bibinfo {author} {\bibfnamefont {S.~S.}\ \bibnamefont
  {Krishtopenko}}, \bibinfo {author} {\bibfnamefont {I.}~\bibnamefont
  {Yahniuk}}, \bibinfo {author} {\bibfnamefont {D.~B.}\ \bibnamefont {But}},
  \bibinfo {author} {\bibfnamefont {V.~I.}\ \bibnamefont {Gavrilenko}},
  \bibinfo {author} {\bibfnamefont {W.}~\bibnamefont {Knap}}, \ and\ \bibinfo
  {author} {\bibfnamefont {F.}~\bibnamefont {Teppe}},\ }\href {\doibase
  10.1103/PhysRevB.94.245402} {\bibfield  {journal} {\bibinfo  {journal} {Phys.
  Rev. B}\ }\textbf {\bibinfo {volume} {94}},\ \bibinfo {pages} {245402}
  (\bibinfo {year} {2016})}\BibitemShut {NoStop}%
\bibitem [{\citenamefont {Vurgaftman}\ \emph {et~al.}(2001)\citenamefont
  {Vurgaftman}, \citenamefont {Meyer},\ and\ \citenamefont
  {Ram-Mohan}}]{SMref3q}%
  \BibitemOpen
  \bibfield  {author} {\bibinfo {author} {\bibfnamefont {I.}~\bibnamefont
  {Vurgaftman}}, \bibinfo {author} {\bibfnamefont {J.~R.}\ \bibnamefont
  {Meyer}}, \ and\ \bibinfo {author} {\bibfnamefont {L.~R.}\ \bibnamefont
  {Ram-Mohan}},\ }\href {\doibase 10.1063/1.1368156} {\bibfield  {journal}
  {\bibinfo  {journal} {J. Appl. Phys.}\ }\textbf {\bibinfo {volume} {89}},\
  \bibinfo {pages} {5815} (\bibinfo {year} {2001})}\BibitemShut {NoStop}%
\end{thebibliography}

\begin{thebibliography}{6}%
\makeatletter
\providecommand \@ifxundefined [1]{%
 \@ifx{#1\undefined}
}%
\providecommand \@ifnum [1]{%
 \ifnum #1\expandafter \@firstoftwo
 \else \expandafter \@secondoftwo
 \fi
}%
\providecommand \@ifx [1]{%
 \ifx #1\expandafter \@firstoftwo
 \else \expandafter \@secondoftwo
 \fi
}%
\providecommand \natexlab [1]{#1}%
\providecommand \enquote  [1]{``#1''}%
\providecommand \bibnamefont  [1]{#1}%
\providecommand \bibfnamefont [1]{#1}%
\providecommand \citenamefont [1]{#1}%
\providecommand \href@noop [0]{\@secondoftwo}%
\providecommand \href [0]{\begingroup \@sanitize@url \@href}%
\providecommand \@href[1]{\@@startlink{#1}\@@href}%
\providecommand \@@href[1]{\endgroup#1\@@endlink}%
\providecommand \@sanitize@url [0]{\catcode `\\12\catcode `\$12\catcode
  `\&12\catcode `\#12\catcode `\^12\catcode `\_12\catcode `\%12\relax}%
\providecommand \@@startlink[1]{}%
\providecommand \@@endlink[0]{}%
\providecommand \url  [0]{\begingroup\@sanitize@url \@url }%
\providecommand \@url [1]{\endgroup\@href {#1}{\urlprefix }}%
\providecommand \urlprefix  [0]{URL }%
\providecommand \Eprint [0]{\href }%
\providecommand \doibase [0]{http://dx.doi.org/}%
\providecommand \selectlanguage [0]{\@gobble}%
\providecommand \bibinfo  [0]{\@secondoftwo}%
\providecommand \bibfield  [0]{\@secondoftwo}%
\providecommand \translation [1]{[#1]}%
\providecommand \BibitemOpen [0]{}%
\providecommand \bibitemStop [0]{}%
\providecommand \bibitemNoStop [0]{.\EOS\space}%
\providecommand \EOS [0]{\spacefactor3000\relax}%
\providecommand \BibitemShut  [1]{\csname bibitem#1\endcsname}%
\let\auto@bib@innerbib\@empty
\bibitem [{\citenamefont {Krishtopenko}\ \emph {et~al.}(2016)\citenamefont
  {Krishtopenko}, \citenamefont {Yahniuk}, \citenamefont {But}, \citenamefont
  {Gavrilenko}, \citenamefont {Knap},\ and\ \citenamefont {Teppe}}]{SMref1}%
  \BibitemOpen
  \bibfield  {author} {\bibinfo {author} {\bibfnamefont {S.~S.}\ \bibnamefont
  {Krishtopenko}}, \bibinfo {author} {\bibfnamefont {I.}~\bibnamefont
  {Yahniuk}}, \bibinfo {author} {\bibfnamefont {D.~B.}\ \bibnamefont {But}},
  \bibinfo {author} {\bibfnamefont {V.~I.}\ \bibnamefont {Gavrilenko}},
  \bibinfo {author} {\bibfnamefont {W.}~\bibnamefont {Knap}}, \ and\ \bibinfo
  {author} {\bibfnamefont {F.}~\bibnamefont {Teppe}},\ }\href {\doibase
  10.1103/PhysRevB.94.245402} {\bibfield  {journal} {\bibinfo  {journal} {Phys.
  Rev. B}\ }\textbf {\bibinfo {volume} {94}},\ \bibinfo {pages} {245402}
  (\bibinfo {year} {2016})}\BibitemShut {NoStop}%
\bibitem [{\citenamefont {Liu}\ \emph {et~al.}(2008)\citenamefont {Liu},
  \citenamefont {Hughes}, \citenamefont {Qi}, \citenamefont {Wang},\ and\
  \citenamefont {Zhang}}]{SMref2}%
  \BibitemOpen
  \bibfield  {author} {\bibinfo {author} {\bibfnamefont {C.}~\bibnamefont
  {Liu}}, \bibinfo {author} {\bibfnamefont {T.~L.}\ \bibnamefont {Hughes}},
  \bibinfo {author} {\bibfnamefont {X.-L.}\ \bibnamefont {Qi}}, \bibinfo
  {author} {\bibfnamefont {K.}~\bibnamefont {Wang}}, \ and\ \bibinfo {author}
  {\bibfnamefont {S.-C.}\ \bibnamefont {Zhang}},\ }\href {\doibase
  10.1103/PhysRevLett.100.236601} {\bibfield  {journal} {\bibinfo  {journal}
  {Phys. Rev. Lett.}\ }\textbf {\bibinfo {volume} {100}},\ \bibinfo {pages}
  {236601} (\bibinfo {year} {2008})}\BibitemShut {NoStop}%
\bibitem [{\citenamefont {Vurgaftman}\ \emph {et~al.}(2001)\citenamefont
  {Vurgaftman}, \citenamefont {Meyer},\ and\ \citenamefont
  {Ram-Mohan}}]{SMref3}%
  \BibitemOpen
  \bibfield  {author} {\bibinfo {author} {\bibfnamefont {I.}~\bibnamefont
  {Vurgaftman}}, \bibinfo {author} {\bibfnamefont {J.~R.}\ \bibnamefont
  {Meyer}}, \ and\ \bibinfo {author} {\bibfnamefont {L.~R.}\ \bibnamefont
  {Ram-Mohan}},\ }\href {\doibase 10.1063/1.1368156} {\bibfield  {journal}
  {\bibinfo  {journal} {J. Appl. Phys.}\ }\textbf {\bibinfo {volume} {89}},\
  \bibinfo {pages} {5815} (\bibinfo {year} {2001})}\BibitemShut {NoStop}%
\bibitem [{\citenamefont {Nichele}\ \emph {et~al.}(2014)\citenamefont
  {Nichele}, \citenamefont {Pal}, \citenamefont {Pietsch}, \citenamefont {Ihn},
  \citenamefont {Ensslin}, \citenamefont {Charpentier},\ and\ \citenamefont
  {Wegscheider}}]{SMref4}%
  \BibitemOpen
  \bibfield  {author} {\bibinfo {author} {\bibfnamefont {F.}~\bibnamefont
  {Nichele}}, \bibinfo {author} {\bibfnamefont {A.~N.}\ \bibnamefont {Pal}},
  \bibinfo {author} {\bibfnamefont {P.}~\bibnamefont {Pietsch}}, \bibinfo
  {author} {\bibfnamefont {T.}~\bibnamefont {Ihn}}, \bibinfo {author}
  {\bibfnamefont {K.}~\bibnamefont {Ensslin}}, \bibinfo {author} {\bibfnamefont
  {C.}~\bibnamefont {Charpentier}}, \ and\ \bibinfo {author} {\bibfnamefont
  {W.}~\bibnamefont {Wegscheider}},\ }\href {\doibase
  10.1103/PhysRevLett.112.036802} {\bibfield  {journal} {\bibinfo  {journal}
  {Phys. Rev. Lett.}\ }\textbf {\bibinfo {volume} {112}},\ \bibinfo {pages}
  {036802} (\bibinfo {year} {2014})}\BibitemShut {NoStop}%
\bibitem [{\citenamefont {Avogadri}\ \emph {et~al.}(2022)\citenamefont
  {Avogadri}, \citenamefont {Gebert}, \citenamefont {Krishtopenko},
  \citenamefont {Castillo}, \citenamefont {Consejo}, \citenamefont {Ruffenach},
  \citenamefont {Roblin}, \citenamefont {Bray}, \citenamefont {Krupko},
  \citenamefont {Juillaguet}, \citenamefont {Contreras}, \citenamefont {Wolf},
  \citenamefont {Hartmann}, \citenamefont {H\"ofling}, \citenamefont
  {Boissier}, \citenamefont {Rodriguez}, \citenamefont {Nanot}, \citenamefont
  {Tourni\'e}, \citenamefont {Teppe},\ and\ \citenamefont {Jouault}}]{SMref5}%
  \BibitemOpen
  \bibfield  {author} {\bibinfo {author} {\bibfnamefont {C.}~\bibnamefont
  {Avogadri}}, \bibinfo {author} {\bibfnamefont {S.}~\bibnamefont {Gebert}},
  \bibinfo {author} {\bibfnamefont {S.~S.}\ \bibnamefont {Krishtopenko}},
  \bibinfo {author} {\bibfnamefont {I.}~\bibnamefont {Castillo}}, \bibinfo
  {author} {\bibfnamefont {C.}~\bibnamefont {Consejo}}, \bibinfo {author}
  {\bibfnamefont {S.}~\bibnamefont {Ruffenach}}, \bibinfo {author}
  {\bibfnamefont {C.}~\bibnamefont {Roblin}}, \bibinfo {author} {\bibfnamefont
  {C.}~\bibnamefont {Bray}}, \bibinfo {author} {\bibfnamefont {Y.}~\bibnamefont
  {Krupko}}, \bibinfo {author} {\bibfnamefont {S.}~\bibnamefont {Juillaguet}},
  \bibinfo {author} {\bibfnamefont {S.}~\bibnamefont {Contreras}}, \bibinfo
  {author} {\bibfnamefont {A.}~\bibnamefont {Wolf}}, \bibinfo {author}
  {\bibfnamefont {F.}~\bibnamefont {Hartmann}}, \bibinfo {author}
  {\bibfnamefont {S.}~\bibnamefont {H\"ofling}}, \bibinfo {author}
  {\bibfnamefont {G.}~\bibnamefont {Boissier}}, \bibinfo {author}
  {\bibfnamefont {J.-B.}\ \bibnamefont {Rodriguez}}, \bibinfo {author}
  {\bibfnamefont {S.}~\bibnamefont {Nanot}}, \bibinfo {author} {\bibfnamefont
  {E.}~\bibnamefont {Tourni\'e}}, \bibinfo {author} {\bibfnamefont
  {F.}~\bibnamefont {Teppe}}, \ and\ \bibinfo {author} {\bibfnamefont
  {B.}~\bibnamefont {Jouault}},\ }\href {\doibase
  10.1103/PhysRevResearch.4.L042042} {\bibfield  {journal} {\bibinfo  {journal}
  {Phys. Rev. Res.}\ }\textbf {\bibinfo {volume} {4}},\ \bibinfo {pages}
  {L042042} (\bibinfo {year} {2022})}\BibitemShut {NoStop}%
\bibitem [{\citenamefont {Nguyen}\ \emph {et~al.}(2016)\citenamefont {Nguyen},
  \citenamefont {Kiselev}, \citenamefont {Noah}, \citenamefont {Yi},
  \citenamefont {Qu}, \citenamefont {Beukman}, \citenamefont {de~Vries},
  \citenamefont {van Veen}, \citenamefont {Nadj-Perge}, \citenamefont
  {Kouwenhoven}, \citenamefont {Kjaergaard}, \citenamefont {Suominen},
  \citenamefont {Nichele}, \citenamefont {Marcus}, \citenamefont {Manfra},\
  and\ \citenamefont {Sokolich}}]{SMref6}%
  \BibitemOpen
  \bibfield  {author} {\bibinfo {author} {\bibfnamefont {B.-M.}\ \bibnamefont
  {Nguyen}}, \bibinfo {author} {\bibfnamefont {A.~A.}\ \bibnamefont {Kiselev}},
  \bibinfo {author} {\bibfnamefont {R.}~\bibnamefont {Noah}}, \bibinfo {author}
  {\bibfnamefont {W.}~\bibnamefont {Yi}}, \bibinfo {author} {\bibfnamefont
  {F.}~\bibnamefont {Qu}}, \bibinfo {author} {\bibfnamefont {A.~J.~A.}\
  \bibnamefont {Beukman}}, \bibinfo {author} {\bibfnamefont {F.~K.}\
  \bibnamefont {de~Vries}}, \bibinfo {author} {\bibfnamefont {J.}~\bibnamefont
  {van Veen}}, \bibinfo {author} {\bibfnamefont {S.}~\bibnamefont
  {Nadj-Perge}}, \bibinfo {author} {\bibfnamefont {L.~P.}\ \bibnamefont
  {Kouwenhoven}}, \bibinfo {author} {\bibfnamefont {M.}~\bibnamefont
  {Kjaergaard}}, \bibinfo {author} {\bibfnamefont {H.~J.}\ \bibnamefont
  {Suominen}}, \bibinfo {author} {\bibfnamefont {F.}~\bibnamefont {Nichele}},
  \bibinfo {author} {\bibfnamefont {C.~M.}\ \bibnamefont {Marcus}}, \bibinfo
  {author} {\bibfnamefont {M.~J.}\ \bibnamefont {Manfra}}, \ and\ \bibinfo
  {author} {\bibfnamefont {M.}~\bibnamefont {Sokolich}},\ }\href {\doibase
  10.1103/PhysRevLett.117.077701} {\bibfield  {journal} {\bibinfo  {journal}
  {Phys. Rev. Lett.}\ }\textbf {\bibinfo {volume} {117}},\ \bibinfo {pages}
  {077701} (\bibinfo {year} {2016})}\BibitemShut {NoStop}%
\end{thebibliography}

%

\end{document}